\documentclass[10pt]{article}

\usepackage[final]{epsfig}
\usepackage{amsmath}
\usepackage{parskip}

\def\lsim{\mathrel{\rlap{\lower4pt\hbox{\hskip1pt$\sim$}}
    \raise1pt\hbox{$<$}}}         
\def\gsim{\mathrel{\rlap{\lower4pt\hbox{\hskip1pt$\sim$}}
    \raise1pt\hbox{$>$}}}         

\def\overleftrightarrow#1{\vbox{\ialign{##\crcr
    $\leftrightarrow$\crcr
    \noalign{\kern 1pt\nointerlineskip}
    $\hfil\displaystyle{#1}\hfil$\crcr}}}

\begin{document}

\hspace{12cm}{\bf TUM/T39-01-16}\\

\begin{center}
{\bf \Large Phases of QCD, Thermal Quasiparticles and\\
 Dilepton Radiation from a Fireball\footnote{work supported in part by BMBF and GSI}}
\end{center}
\begin{center}
{Thorsten Renk$^{a}$, Roland Schneider$^{a}$, Wolfram Weise$^{a,b}$}

{\small \em $^{a}$ Physik Department, Technische Universit\"{a}t M\"{u}nchen,
D-85747 Garching, GERMANY}

{\small \em$^{b}$ ECT$^\ast$, Villa Tambosi, I-38050 Villazzano (Trento), ITALY}
\end{center}
\vspace{0.25 in}

\begin{abstract}
We calculate dilepton production rates from a fireball
adapted to the kinematical conditions realized
in ultrarelativistic heavy ion collisions over a broad range of beam energies.
The freeze-out state of the fireball is fixed by hadronic observables.
We use this information combined with the initial geometry of the collision region
to follow the space-time evolution of the fireball. Assuming entropy conservation,
its bulk thermodynamic properties can then be uniquely obtained once the equation of 
state (EoS) is specified.
The high-temperature (QGP) phase is modelled by a non-perturbative quasiparticle
model that incorporates a phenomenological confinement
description, adapted to lattice QCD results. 
For the hadronic phase, we interpolate the EoS into the region where
a resonance gas approach seems applicable, keeping track of a possible
overpopulation of the pion phase space. In this way, the fireball evolution
is specified without reference to dilepton data, thus
eliminating it as an adjustable parameter in the rate calculations.
Dilepton emission in the QGP phase is then calculated within the quasiparticle model.
In the hadronic phase, both temperature and finite baryon density effects on the photon
spectral function are incorporated. Existing dilepton data from CERES at 158 and 40 AGeV Pb-Au
collisions are well described, and a prediction for the PHENIX
setup at RHIC for $\sqrt{s}$ = 200 AGeV is given.
\end{abstract}

\vspace {0.25 in}

\section{Introduction}
\label{sec_introduction}

There are convincing arguments from lattice simulations \cite{lat1} that  QCD
undergoes a phase transition at a temperature of about 150 - 170 
MeV \cite{lat2} from a confined hadronic phase to a
phase where quarks and gluons constitute the relevant degrees of freedom, the quark-gluon plasma (QGP).
It is believed that chiral symmetry restoration, associated with the vanishing of the chiral condensate
$\langle \bar{q} q \rangle$, takes place at about the same temperature. This new state of matter is
clearly of enormous interest, and the search for experimental evidence of its creation in
ultrarelativistic heavy ion collisions (URHIC) is going on. However, it is still a matter of debate
whether
the QGP has already been encountered. Any signature of the QGP is folded with the
time evolution (and consequently the
evolution of volume, temperature and baryon density) of the fireball created in such a collision.
Furthermore, this evolution continues
after the phase transition and any information is thus mixed with signals from the hadronic
phase. Therefore,
extracting an unambiguous signal for the QGP phase is not an easy task.

Dileptons ($e^+e^-$ and $\mu^+ \mu^-$ pairs) are interesting probes in this context since they do not
interact strongly and escape unthermalized from the hot region at all stages of the collision.
Therefore, in contrast to hadronic probes which can tell only of the late stages (the
freeze-out), dileptons carry information also on the early moments of the collision where the QGP phase
is expected to exist. In the QGP phase, dileptons originate mainly from $q\bar{q}$ annihilation
processes, whereas in the hadronic phase the main sources are pion and kaon annihilation processes which
are enhanced due to the formation of the light vector mesons $\rho, \omega$ and $\phi$. In order to
interpret signals from the QGP, one must obviously first understand the hadronic phase.

Unfortunately, the initial conditions of a fireball formed in
URHIC are not well under control, nor is
its subsequent evolution known in sufficient detail so far.
Different scenarios (pure hadronic, pure QGP and mixed phases)
can all be designed such that they are in rough agreement with experimental data.
This is shown in calculations using microscopic transport equations \cite {EVENT0,EVENT},
hydrodynamical descriptions \cite{HYDRO-SOLLFRANK, HYDRO-SHU}
and simplified approaches \cite{KKW96,RW}.
In this paper, we make attempts to tighten the constraints
on the QGP equation of state (EoS) and the fireball evolution
dynamics by establishing
contacts with lattice QCD thermodynamics and hadronic observables.
We extend the work of \cite{SW00} (and other previous work)
in several directions.

First, we develop a thermodynamically consistent quasiparticle picture,
incorporating important aspects of confinement, to describe both
the fireball dynamics and the dilepton radiation from
the quark-gluon phase at the early stages of the collision.
Perturbation theory is known not to work in
the quark-gluon phase \cite{FAIL} even at temperatures well
above $T_C$. We adopt instead an approach which treats quarks
and gluons as massive thermal quasiparticles, with
their properties determined to reproduce lattice QCD
results.
The driving force of the phase transition, the confinement/deconfinement
process, is given a statistical meaning in terms of a suppression of thermally active
partonic degrees of freedom as the critical temperature is approached from above.
For the hadronic phase, we calculate the
dependence of the photon spectral function (that enters the dilepton rate) 
on temperature and baryon  density using Vector Meson 
Dominance combined with Chiral Dynamics.

Second, a fireball model reminiscent of hydrodynamics is set up. 
The endpoint of the fireball evolution is fixed by requiring agreement
with hadronic observables, such as Hanbury-Brown-Twiss (HBT) radii, $m_t$-spectra and 
$dN/dy$ distributions, which carry information on the freeze-out stage. 
Information on the initial state can be gained by estimates
of the nuclear overlap geometry. We then construct the evolution between
initial and freeze-out conditions
by following the fireball thermodynamics along volume slices of constant proper
time $\tau$. 
This is done under the assumption of entropy conservation where the total
entropy of the system is determined at freeze-out. The EoS in the
QGP phase is calculated using a quasiparticle picture. In the hadronic
phase a smooth interpolation to a hadronic resonance
gas approach is applied. We thus obtain the temperature $T$ of the system
at a given proper time.
This procedure results in a thermodynamically self-consistent description
of the fireball evolution, which is fixed by hadronic observables
only. 

Using this model as an input, we calculate dilepton rates and
compare results to existing CERES/NA45 data of 40 and 160 AGeV 
Pb-Au collisions, finding good agreement in both cases.
We also give predictions for the kinematical conditions at RHIC.
Differences between the present approach and other
works are discussed.

\section{Dileptons from a fireball}
\label{sec_dileptons}

The lepton pair emission rate from a hot domain populated by particles in
thermal equilibrium at temperature $T$ is proportional to the imaginary part
of the spin-averaged, photon self-energy, with these particles as
intermediate states. The thermally excited particles annihilate to yield a
time-like virtual photon with four-momentum $q$ which decays subsequently into
a lepton-antilepton pair. The differential pair production rate is given by
\begin{equation}
\frac{dN}{d^4 x d^4q}  =  \frac{\alpha^2}{\pi^3 q^2} \ \frac{1}{e^{\beta q^0}
- 1} \ \mbox{Im}\bar{\Pi}(q, T) = \frac{\alpha^2}{12\pi^4} \frac{R(q, T)}{e^{\beta q^0}
- 1}  ,\label{dileptonrates}
\end{equation}
where $\alpha = e^2/4\pi$, $\beta = 1/T$, and we have neglected the lepton
masses. We have defined $\bar{\Pi}(q) = -\Pi^\mu_{\ \mu} /3$ and introduce the averaged photon spectral function $R(q, T) = (12\pi/q^2) \ \mbox{Im}\bar{\Pi}(q,T)$. Here $\Pi^\mu_{\
\mu}$ denotes the trace over the thermal photon self-energy which is
equivalent to the thermal current-current correlation function
\begin{equation}
\Pi_{\mu\nu}(q,T) = i \int d^4 x \ e^{iqx} \langle \mathcal{T} j_{\mu}(x)
j_{\nu}(0) \rangle_\beta,
\end{equation}
where $j_\mu$ is the electromagnetic current.
Eq.(\ref{dileptonrates}) is valid to order $\alpha$ in the electromagnetic
interaction and to all orders in the strong interaction.

In order to compare with experimental data, we construct a model for the space-time
evolution of a heavy-ion collision, assuming approximate thermal equilibrium
to be a useful concept \cite{EQ1}. Some recent discussions suggest
\cite{EQ2,EQ3} that equilibration times at the conditions of heavy ion
collisions at SPS and RHIC may indeed be very short (possibly less than 1 fm/c), small
compared to expansion times of order 10 - 20 fm/c, although this is still a
matter of debate. The proper approach for our purposes is to use a
simplified fireball model which employs a hot cylinder, expanding
isentropically both in the longitudinal and transverse direction.
Temperature and density are assumed to be spatially homogeneous at a given proper 
time. In chapter \ref{sec_fireball}, this fireball model is explained in detail.

The differential rate of eq.(\ref{dileptonrates}) is integrated over the space-time history of the collision to
compare the calculated dilepton rates with the CERES/NA45 data \cite{CERES}
taken in Pb-Au collisions at 160 AGeV (corresponding to
a c.m. energy of $\sqrt{s} \sim 17$ AGeV) and 40 AGeV ($\sqrt{s} \sim 8$ AGeV).  The CERES experiment is a
fixed-target experiment. In the lab frame, the CERES
detector covers the limited rapidity interval $\eta = 2.1-2.65$, {\em i.e.} $\Delta\eta = 0.55$. 
We integrate the calculated rates over the
transverse momentum $p_T$ and average over $\eta$, given that $d^4p = M p_T \
dM  \ d\eta \ dp_T \ d\theta.$
The formula for the space-time- and $p$-integrated dilepton rates hence becomes
\begin{equation}
\frac{d^2N}{dM d\eta} =  \frac{2\pi M}{\Delta \eta} \int \limits_0^{\tau_{f}}
d\tau \  \int  d\eta \ V(\eta,T(\tau))\int
\limits_0^\infty dp_T \ p_T
 \ \frac{dN(T(\tau),M, \eta,
p_T)}{d^4 x d^4p} \ Acc(M, \eta, p_T), \label{integratedrates}
\end{equation}
where $\tau_{f}$ is the freeze-out proper time of
the collision, $V(\eta,T(\tau))$ describes the proper time evolution of 
volume elements moving at different rapidities 
and the function $Acc(M, \eta, p_T)$ accounts for the
experimental acceptance cuts specific to the detector. At the CERES
experiment, each electron/positron track is required to have a transverse
momentum $p_T > 0.2$ GeV, to fall into the rapidity interval $2.1 < \eta <
2.65$ in the lab frame and to have a pair opening angle $\Theta_{ee} > 35$
mrad. Eq.(\ref{integratedrates}) is then convoluted with the finite energy resolution of the detector. 
Finally, for comparison with the CERES data, the resulting rate 
is divided by $dN_{ch}/d\eta$, the rapidity density of charged particles.

RHIC operates as a collider experiment, so in this case the 
fireball is centered around $\eta = 0$.
Here, the PHENIX detector acceptance can be schematically modelled by requiring
that each electron/positron track falls
in the rapidity interval $-0.35 < \eta <0.35$, has transverse
momentum $p_T> 0.2$ GeV and a pair opening angle
of $\Theta_{e\overline{e}} > 35$ mrad. At present,
an abundance of data on Au-Au collisions at $\sqrt{s}$ = 130 AGeV have already been analysed, and first data of the run at the higher energy $\sqrt{s} = $ 200 AGeV are available.
\section{Quasiparticle description of the quark-gluon phase}
\label{sec_quasiparticles}
In order to proceed, we need to develop a picture of
the QCD dynamics in the quark-gluon plasma
phase close to the phase transition where the system presumably spends an
appreciable fraction of its time
in URHIC at current kinematical conditions.
A realistic equation of state is needed to model the expanding fireball, and
the calculation of the differential dilepton emission rate requires the
photon self-energy with partonic intermediate states.

Since it is expected that $\alpha_s(T) \rightarrow 0$ at very high temperatures,  
one might naively expect that a perturbative
approach to calculate these quantities is possible.
However, it is now well-known that perturbation theory
breaks down for gauge theories at finite temperature because of the occurrence
of infrared divergences \cite{FAIL}. The
Hard Thermal Loop (HTL) calculation scheme consistently resums all contributions to
a given order in the gauge coupling constant $g_s$ and is explicitly gauge
invariant. Its application relies on a scale separation into 'hard'
($p \sim T$) and 'soft' ($p \sim g_s T$) momentum modes with $g_s \ll
1$, though for QCD, this condition is likely to be fulfilled only at extremely
large $T$
far outside the scope of present and future experiments \cite{LOW_G}. In addition,
non-abelian gauge theories also suffer from a dependence on an 'ultra-soft'
magnetic mass $m_{mag} \sim g_s^2 T$ which is intrinsically nonperturbative.

The imaginary part of the photon self-energy Im$\bar{\Pi}$ appearing in
eq.(\ref{dileptonrates}) has been calculated to two-loop order in thermal
perturbation theory \cite{AG98}, and there exist estimates for the
three-loop contributions \cite{AG00}. So far, the convergence of the resulting series is
badly behaved even for small couplings $g_s$. A similar behaviour has been
found in the perturbative calculation of the free energy of a QGP and also in
simple scalar theory. Furthermore, close to $T_C$ we expect intrinsically
non-perturbative dynamics to enter: the confinement/deconfinement transition
and chiral symmetry breaking are not amenable in an expansion in $g_s$.
In the light of these facts, we will use a more phenomenological approach to
QCD thermodynamics in this work.

One way to obtain nonperturbative input is through finite temperature lattice
simulations of a pure gluonic or quark-gluon plasma. We have shown
recently that it is possible to describe the equation of state (EoS) of such
systems to a very good approximation by the EoS of a  gas of
quasiparticles with thermally generated masses, incorporating confinement
phenomenologically by a temperature-dependent effective number of active degrees of
freedom. Here we give a short summary of the method and refer the reader to
\cite{RW01} for a more detailed discussion.

From asymptotic freedom, we expect that at extremely  high temperatures the plasma
consists of quasifree quarks and gluons. HTL perturbative calculations
find, for thermal momenta, spectral functions of the form
$\delta(E^2 - k^2 - m^2(T))$ with $m(T) \sim g_s T$.
As long as the spectral functions of
the thermal excitations at lower temperatures resemble qualitatively this
asymptotic form, a quasiparticle description is expected to be
applicable. The QCD dynamics is then incorporated in the thermal
masses of the 'effective' quarks and gluons and in a function $B(T)$
which is necessary for thermodynamical consistency and plays the r\^{o}le of
the thermal vacuum energy.

Since the existence of a preferred frame of reference
(in which the hot matter is at rest) breaks Lorentz
invariance, new partonic excitations (longitudinal gluonic plasmons and
helicity-flipped plasminos) are also present in the plasma. However, their
spectral strengths are exponentially suppressed for high temperatures and
momenta $k \sim T$ which dominate the integrals of macroscopic
thermodynamic quantities such as pressure, entropy and energy density, therefore  their contributions can be neglected. This is also in line with results from a lattice calculation of the quark and gluon propagators \cite{LAT_COLLECT}. For the differential dilepton rate, their
influence is not so easily discarded, and we will discuss that in section \ref{sec_calculation}.

The thermal excitations can then be described by a dispersion equation
\begin{equation}
E^2(k,T) = k^2 + m^2_i(T).
\end{equation}
Here, $k = |\mathbf{k}|$, and the subscript $i$ labels the particle species: $i = g$ for gluons
and $i = q$ for quarks. $m_i(T)$ stands for a thermal mass which is derived
from the self-energy of the corresponding particle, evaluated at thermal
momenta $E, k \sim T$. This is expected to be meaningful as long
as the self-energy is only weakly momentum dependent in that kinematic region.
Additionally, for a quasiparticle to be a meaningful concept at all, we require the
imaginary part of the self-energy, and hence its thermal width, to be small.

The gluon mass, following ref.~\cite{RW01}, becomes
\begin{equation}
\frac{m_g(T)}{T} = \sqrt{\frac{N_C}{6} + \frac{N_f}{12}} \ \tilde{g}(T, N_C,
N_f) \label{m_gluon} %
\end{equation}
with the effective coupling specified as
\begin{equation}
\tilde{g}(T, N_C, N_f) = \frac{g_0}{\sqrt{11N_C - 2N_f}} \left( [1 + \delta] - \frac{T_C}{T}
\right)^{\gamma}. \label{g_tilde}
\end{equation}
$N_C$ and $N_f$ stand for the number of colours and flavours, respectively.
The functional dependence of $m_g(T)$ on $T$ is based on the conjecture that
the phase transition is weakly first order or second order which suggests an almost
powerlike behaviour $m \sim (T - T_C)^\gamma$ with some pseudo-critical
exponent $\gamma > 0$. Setting $g_0 = 9.4$, $\delta = 10^{-6}$ and $\gamma =
0.1$, the effective mass (\ref{m_gluon}) approaches the HTL result at high temperatures.

The thermal quark mass reads
\begin{equation}
\frac{m_q(T)}{T} = \sqrt{ \left(\frac{m_{q,0}}{T} + \frac{1}{4}\sqrt{\frac{N_C^2-1}{N_C}} \tilde{g}(T) \right)^2 +
\frac{N_C^2-1}{16 N_C} \tilde{g}(T)^2  } \label{m_quark}
\end{equation}
with the zero-temperature bare quark mass $m_{q,0}$. Note that in contrast to
previous quasiparticle models extrapolated from HTL calculations \cite{PKS00, LU98}, the thermal
masses used here {\em drop} as $T_C$ is approached from above. Of course, for
$T\gg T_C$, the near-critical behaviour inferred in (\ref{g_tilde}) ceases to be valid
and the perturbative limit of $m_g(T)$ and $m_q(T)$ will be recovered.

\begin{figure}[!hbt]
\begin{center}
\epsfig{file=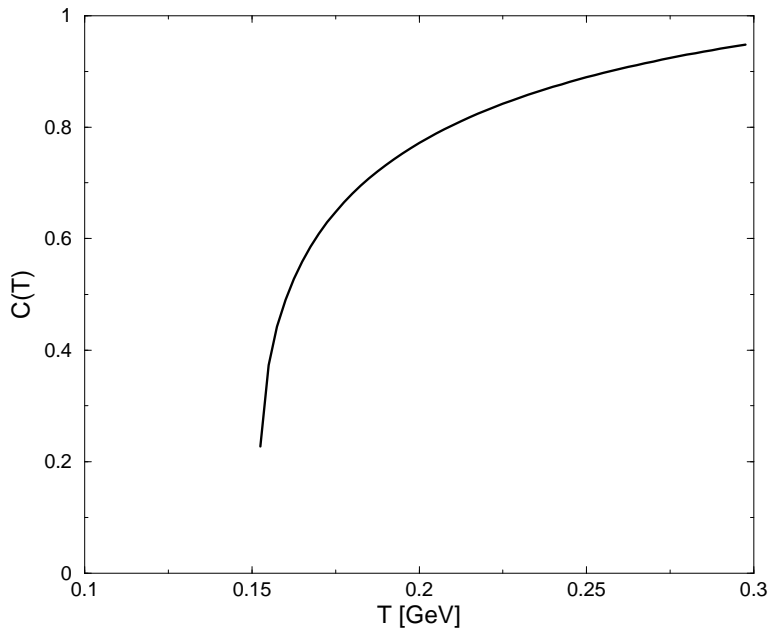,width=8.5cm, angle = -90}
\end{center}
\caption{The confinement factor $C(T)$ as a function of temperature for two
light flavours ($m_{u,d} = 0$) and one heavier flavour ($m_s \simeq 170$ MeV).}
\label{C_T}
\end{figure}

The pressure of
the QGP system takes the following form:
\begin{equation}
p(T) = \frac{\nu_g}{6\pi^2} \int_0^\infty dk \left[ C(T)f_B(E_k^g) \right] \frac{k^4}{E_k^g} + \sum
\limits_{i=1}^{N_f}  \frac{2 N_C}{3\pi^2} \int_0^\infty dk \left[ C(T)f_D(E_k^i) \right]
\frac{k^4}{E_k^i} - B(T). \label{p3_def}
\end{equation}
Energy density $\epsilon(T)$ and entropy density $s(T)$ follow accordingly
(see \cite{RW01} for details):

\begin{equation}
\epsilon(T) = \frac{\nu_g}{2\pi^2} \int_0^\infty dk k^2 [C(T)f_B(E^g_k)] E^g_k
+ \sum\limits_{i=1}^{N_f} \frac{2 N_C}{\pi^2} \int_0^\infty dk k^2 [C(T)f_D(E_k^i)] E_k^i+ B(T),
\end{equation}

\begin{equation}
s(T) = \frac{\nu_g}{2\pi^2T} \int_0^\infty dk k^2 [C(T) f_B(E^g_k)] \frac{\frac{4}{3}k^2 + m_g^2(T)}{E_k^g}
+ \sum\limits_{i=1}^{N_f} \frac{2 N_C}{\pi^2 T} \int_0^\infty dk k^2[C(T)f_D(E_k^i)]
\frac{\frac{4}{3}k^2 + m_i^2(T)}{E_k^i}.
\end{equation}
Here, $E_k^g = \sqrt{k^2 + m_g^2(T)}$ and $E_k^i = \sqrt{k^2 + m_i^2(T)}$ for each quark
flavour $q = i$. The multiplicity $\nu_g = 16$ counts the
transverse gluonic degrees of freedom.
The bosonic particle distribution function reads $f_B(E) = (\exp(E/T) -
1)^{-1}$, the fermionic one is $f_D(E) = (\exp(E/T) + 1)^{-1}$.

The function $C(T)$ accounts, in a statistical sense, for the onset of
confinement by reducing the number of thermally active degrees of freedom.
Its explicit form is obtained by calculating the entropy density
of the QGP with the gluon mass (\ref{m_gluon}) and the quark mass
(\ref{m_quark}). Dividing the lattice entropy density by this result yields
$C(T)$. It can be parametrized as
\begin{equation}
C(T) = C_0 \left( [1 + \delta_c] - \frac{T_C}{T} \right)^{\gamma_c}.
\end{equation}
For two light quarks and one heavy quark, the parameters take the values $C_0 = 1.16$,
$\delta_c = 0.02$ and $\gamma_c = 0.29$ (see figure \ref{C_T})\footnote{We
mention that the proposed method relies on input from the lattice. Whereas high
precision data exist in the pure gauge sector, calculations with dynamical
quarks are not yet in a satisfactory position to yield proper
continuum-extrapolated results with physical quark masses. We estimate that
the results obtained within our confinement model may still change by 5-15\% in
the vicinity of $T_C$ once high statistics data are available. However, this
small correction does not influence the results of the following discussion.}.
The function $B(T)$ is now uniquely determined from $m_i(T)$
and $C(T)$ up to an integration constant that is fixed by Gibbs' condition
$p_{QGP} = p_{hadr}$ at $T_C$.

\begin{figure}[htb]
\begin{center}
\epsfig{file=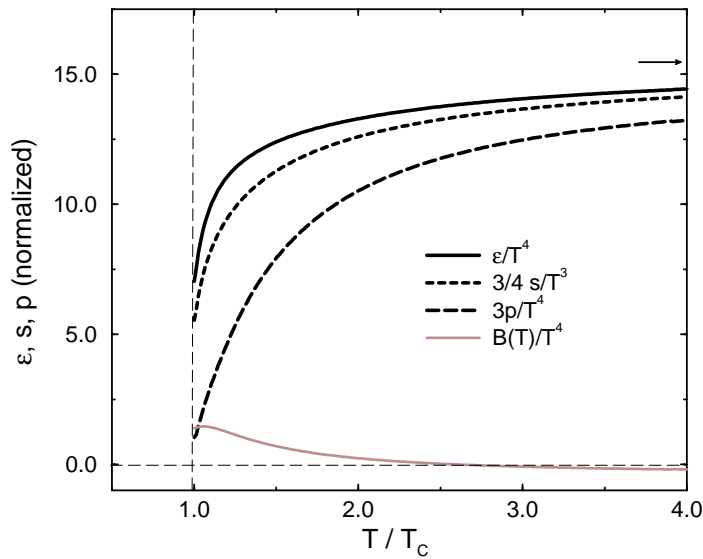,width=8.5cm, angle = -90}
\end{center}
\caption{Pressure, energy density and entropy density for two
light quark flavours ($m_{u,d} = 0$) and a heavier strange quark
($m_s \simeq 170 $ MeV) in our quasiparticle model.
The arrow indicates the ideal gas limit of massless
three-flavour QCD. The function $B(T)$ is
also shown.}
\label{2_1_flav}
\end{figure}
Figure \ref{2_1_flav} shows the pressure, energy and entropy density for two
light quark flavours ($m_{u,d} = 0$) and a heavier strange quark ($m_s \simeq
170 $ MeV) in our confinement model.

We now proceed as follows: We adopt the
picture of the QGP as a quasiparticle gas consisting of massive 'quarks' and
'gluons' which are the relevant thermodynamical degrees of freedom. The
resulting EoS serves as input for the construction of the fireball which is
discussed in section \ref{sec_fireball}. Furthermore,
within this framework the quasiparticle $q\bar{q}$-loop becomes the only
contribution to the photon self-energy in eq.(\ref{dileptonrates}) since the
quasiparticles are by construction non-interacting. The
reduced probability for thermally exciting a quark because of
confinement, and the thermal effective masses, have to be taken into account properly
as discussed in the next section.

\section{Calculation of the photon spectral function}
\label{sec_calculation}

\subsection{The quark-gluon phase}

\label{subsec_qgp}

As long as the thermodynamically active degrees of freedom are quarks and
gluons, the timelike photon couples to the continuum of thermally
excited $q\overline{q}$ states and subsequently converts into a charged lepton pair.
The calculation of the photon spectral function  at the one-loop level is performed using
standard thermal field theory methods. The well-known leading-order result
for bare quarks and gluons as degrees of freedom is:
\begin{equation}
\label{E-ImPi}
\begin{split}
\text{Im}\overline{\Pi}(q^0, {\bf q},T) &= -\frac{q^2}{12\pi} \cdot 3 \sum_{f=u,d,s} \theta(q^2-4m_f^2) e_f^2
\left(1+\frac{2m_f^2}{q^2}\right) \sqrt{1-\frac{4m_f^2}{q^2}}\\
\times&\left(1+2\left[\frac{T}{|{\bf q}|} \frac{1}{\sqrt{1-\frac{4m_f^2}{q^2}}}
\ln \left(\frac{f_D\left(\frac{q_0}{2} -\frac{|{\bf q}|}{2} \sqrt{1-\frac{4m_f^2}{q^2}}\right)}
{f_D\left(\frac{q_0}{2} +\frac{|{\bf q}|}{2} \sqrt{1-\frac{4m_f^2}{q^2}}\right)}\right) -1\right]
\right),
\end{split}
\end{equation}
where $q = (q^0, {\bf q})$ is the four-momentum of the virtual photon, $e_f$ the quark electric
charge and $m_f$ the quark mass of flavour $f$. This result, however, holds only
up to perturbative higher order corrections in $g_s$ that take into account
collective plasma effects.
Here, contributions from soft gluons
lead to strong modifications.
The corresponding two- and three-loop
contributions show no clear convergence \cite{AG98, AG00}.
Close to the phase transition, we also expect non-perturbative confinement
physics to enter. Consequently, we follow a different approach.

Recalling the results of the previous section, the thermodynamic properties of the QGP
as given by lattice QCD are well reproduced by a
gas of quasiparticles. Let us now assume that a quark quasiparticle couples to a photon
in the same way as a bare quark (a form factor representing the
'cloud' of the quasiparticle could in principle
also be included, but in absence of information about the detailed quasiparticle
structure we ignore this point). For a gas of non-interacting
quasiparticles, the one-loop result for Im$\overline{\Pi}$ is
adequate, with input properly adjusted.
All higher order QCD effects manifest
in the thermal quasiparticle masses $m(T)$, the function
$B(T)$ and the confinement factor $C(T)$.
Incorporation of the first two features in the calculation is straightforward.
The bare
quark masses in eq.~(\ref{E-ImPi}) simply have to be replaced by the $T$-dependent quasiparticle
masses for each flavour, see eq.~(\ref{m_quark}).
The thermal vacuum energy $B(T)$ does not contribute to the dilepton rate.

The naive replacement $f_D \rightarrow C(T) f_D$ is, however,
not permitted in eq.~(\ref{E-ImPi}). Since any modification of the
free particle distribution functions leads to non-equilibrium field theory, products of
delta functions
(pinch singularities) may arise in loop calculations. Therefore, the quasiparticle model
as it stands cannot be used in
expressions derived from simple perturbative thermal field theory. 
Recalling the physical interpretation of the confinement factor $C(T)$, we can use
the expression for the dilepton rate, eq.~(\ref{dileptonrates}),
instead.  The mechanism for dilepton production at tree-level
is the annihilation of a $q\overline{q}$ pair into a virtual photon where the
quark lines are multiplied by the distributions $f_D(T)$, giving
the probability of finding a quark or an antiquark in the hot medium.
This also becomes clear when we look at the limit ${\bf q} \rightarrow 0$ of eq.~(\ref{E-ImPi}). Then,
\begin{equation}
\mbox{Im}\overline{\Pi}(q^0, T) \sim \mbox{Im}\overline{\Pi}(q^0, T=0) \cdot (1 - 2 f_D(q^0/2)),
\end{equation}
and the temperature enters only in the Pauli-blocking of the quarks propagating 
in the loop. Now, from eq.~(\ref{dileptonrates})
\begin{equation}
\frac{dN}{d^4 x d^4q} \sim  f_B(q^0) \mbox{Im}\bar{\Pi}(q, T).
\end{equation}
Combining the different thermal occupation factors, we end up with the well-known result
\begin{equation}
\frac{dN}{d^4 x d^4q} \sim \left[ f_D(q^0/2) \right]^2,
\end{equation}
so the differential dilepton rate is proportional to the probability of finding a quark $q$ times the probability of finding an antiquark $\bar{q}$ with the correct momentum, as anticipated\footnote{We neglect a possible chemical potential for the quarks. For a finite $\mu$, the corresponding expression would be $\frac{dN}{d^4 x d^4q} \sim f_D((q^0 - \mu)/2) f_D((q^0 + \mu)/2)$.}. The incorporation of the confinement factor is now obvious: since it reduces the number of thermally active degrees of freedom, it also reduces the dilepton rate by a factor of $C(T)^2$.

\begin{figure}[tb]
\begin{center}
\epsfig{file=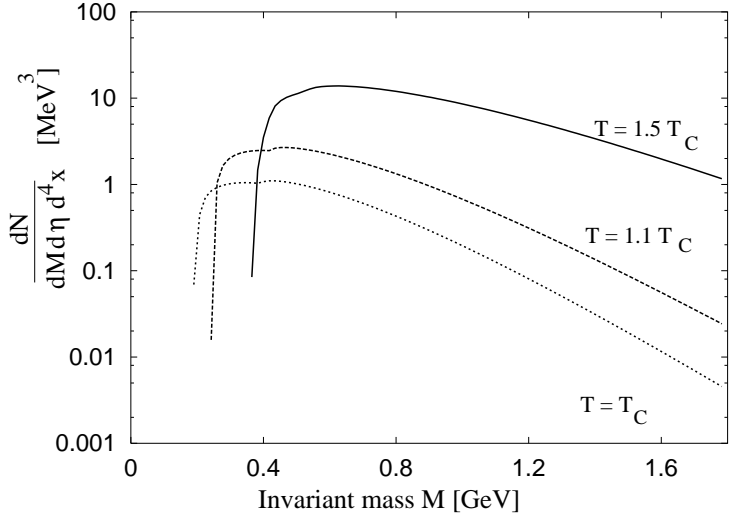,width=10cm}
\end{center}
\caption{\label{F-conf-rate}Dilepton rates originating from the QGP
phase in the quasiparticle model for different temperatures.}
\end{figure}

In summary, eq.~(\ref{dileptonrates}) can be used to calculate the dilepton rate
originating from a hot QGP phase, provided an overall factor $C(T)^2$ is applied to
account for the reduced probabilities, and the bare masses $m_f$
in the one-loop expression (\ref{E-ImPi}) are
replaced by the $T$-dependent thermal masses (\ref{m_quark}).
The r\^ole of the factor $C(T)$ is illustrated in figure~\ref{F-conf-rate},
where the differential dilepton rate originating from a hot QGP in the
quasiparticle approach is shown for different temperatures.
Note that the plotted quantity is independent of the fireball volume,
so the resulting differences are only due to the dropping
quasiparticle masses and the squared confinement factor $C(T)$, which
is responsible for a decrease by more than an order of magnitude at
$T \sim T_C$ as compared to the highest temperature shown.
One also observes that, as expected, the (negative) slope
of the production rate in the region of high invariant mass
gets steeper as the temperature decreases.
It is important to note that this setup neglects
contributions from hadronic degrees of freedom above $T_C$. As mentioned, 
quarks and gluons become clustered into hadrons (glueballs, mesons) as the temperature
approaches $T_C$ from above. These hadronic excitations are comparatively heavy 
and thus do not contribute much to the thermodynamics. 
Since we do not know in detail how the statistical re-arrangement of 
degrees of freedom occurs, we refrain from including these hadronic sources 
of dilepton yield above $T_C$. 
Our calculation is therefore expected to give a {\em lower limit} 
on the leptonic radiation from the QGP phase.

The quasiparticle model does not take into account the collective plasma
modes, such as the (longitudinal) gluonic plasmons and the
(helicity-flipped) quark plasminos. Since their residues are
exponentially suppressed in the HTL approximation for thermal momenta $k
\sim T$, their contributions to the thermodynamical quantities are
negligible. However, in the case of soft dilepton production it is
well known that these modes lead to sharp, distinct structures in the
spectrum, referred to as Van Hove singularities \cite{BPY90}. The plasmino
branch has a minimum in its dispersion relation at $ k \neq 0$ (which
follows on very general grounds \cite{PT00}, independent of the HTL
approximation). This leads to a diverging density of states which, in
turn, shows up in the dilepton spectrum as a pronounced peak. Our model
cannot exhibit, by construction, such plasmino effects. However, since
the peaks are roughly located at $\simeq 2 m_q(T)$, where $m_q(T)$ is the
thermal mass of the quasiparticles, these Van Hove singularities would
be smeared out by the fireball evolution. The thermal mass drops as the
temperature goes down, dragging along the peak position with it.
Furthermore, since $m_q(T)$ is of the order of the temperature $T$ or smaller
in our model, the singularities appear at low invariant mass ($< 500$ MeV)
where they are overwhelmed by the hadronic part of the dilepton
production. Therefore the presence of these collective modes would
presumably not influence our results for the dilepton rate in the QGP phase.

Our model of the QGP phase also finds preliminary support from 
a recent lattice calculation \cite{DILEPTON_LATTICE} of thermal vector 
meson correlation functions above $T_C$ in quenched QCD with 
Clover improved Wilson fermions. Using the maximum entropy method, the vector 
spectral function was extracted from the corresponding current correlator. Although the 
statistical uncertainties are 
still considerable, it is interesting to note that the resulting spectrum 
resembles the free spectral function, as in our case,  
and has a gap at low energies given by a thermal mass threshold of $(2 - 3)T$, 
which is indeed close to $2 m_q(T)$, the natural cut-off of the spectrum 
and, correspondingly, of the thermal dilepton radiation in the confinement 
model (see figure \ref{F-conf-rate}). Furthermore, this result seems to 
rule out heavy quark quasiparticles in the deconfined phase, as predicted 
by other phenomenological models \cite{PKS00, LU98}. Of course, higher 
statistics and improved actions are mandatory to confirm these observations.

\subsection{The hadronic phase}

Below $T_C$, confinement sets in and the effective degrees of freedom
change to colour singlet, bound $q\bar{q}$ or $qqq$
($\bar{q}\bar{q}\bar{q}$) states. The photon couples now to the lowest-lying
'dipole' excitations of the vacuum, the hadronic $J^P = 1^-$ states: the
$\rho$, $\omega$ and $\phi$ mesons and multi-pion states carrying the  same
quantum numbers. The electromagnetic current-current correlation function can
be connected to the currents generated by these mesons using an effective
Lagrangian which approximates the $SU(3)$ flavour sector of QCD at low
energies. The appropriate model for our purposes is the {\em improved Vector
Meson  Dominance} model combined with chiral dynamics of pions and kaons as
described in \cite{KKW1}. Within  this model, the following relation between
the imaginary part of the irreducible photon self-energy  $\mbox{Im}
\bar{\Pi}$ and the vector meson self-energies $\Pi_V(q)$ in vacuum is derived:
\begin{equation}
\mbox{Im} \bar{\Pi}(q) = \sum \limits_V \frac{\mbox{Im}
\Pi_V(q)}{g_V^2} \ |F_V(q)|^2, \label{ImBarPi}  \quad F_V(q) = \frac{\left( 1-
\frac{g}{g^0_{V}} \right)q^2 - m_V^2}{q^2 - m_V^2 + i  \mbox{Im}\Pi_V(q)},
\end{equation}
where $m_V$ are the (renormalized) vector meson masses,
$g^0_V$ is the $\gamma V$ coupling and $g$ is the vector meson coupling
to the pseudoscalar Goldstone bosons $\pi^\pm, \pi^0$
and $K^\pm, K^0$. Eq.(\ref{ImBarPi}) is valid for a virtual photon with
vanishing three-momentum $\mathbf{q}$. For finite three-momenta there exist two
scalar functions $\bar{\Pi}_L$ and $\bar{\Pi}_T$, because the existence of a
preferred frame of reference (the heat bath) breaks Lorentz invariance, and
one has to properly average over them. However, taking the limit $|\mathbf{q}|
\rightarrow 0$ should be reasonable for our purposes in view of the fact that
the c.m. rapidity interval accessible at CERES and RHIC restricts $|\mathbf{q}|$ on
average to only a fraction of the vector meson mass $m_V$.

\begin{figure}[h!t]
\begin{center}
\epsfig{file=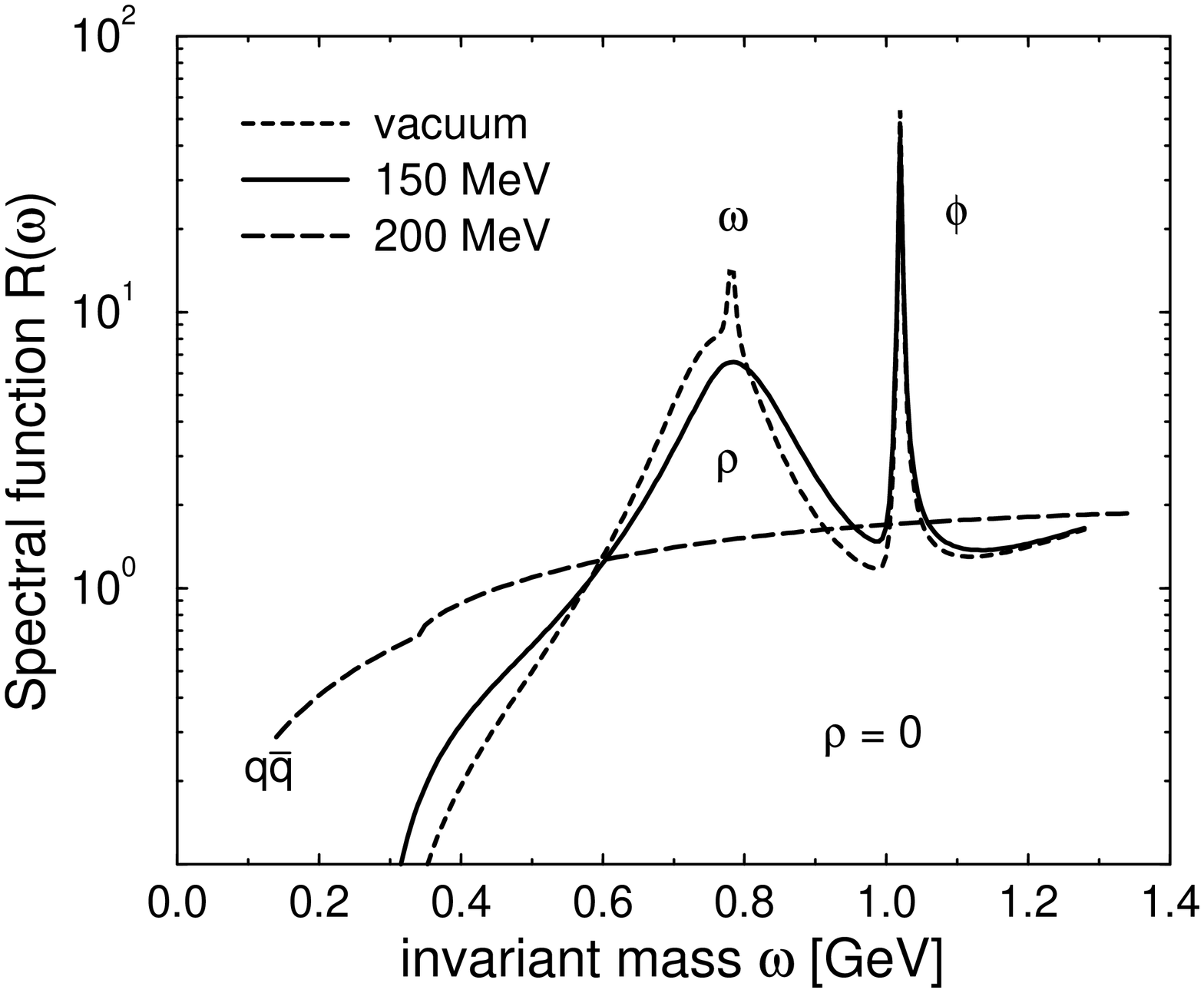,width=6.0cm, angle = -90}
\epsfig{file=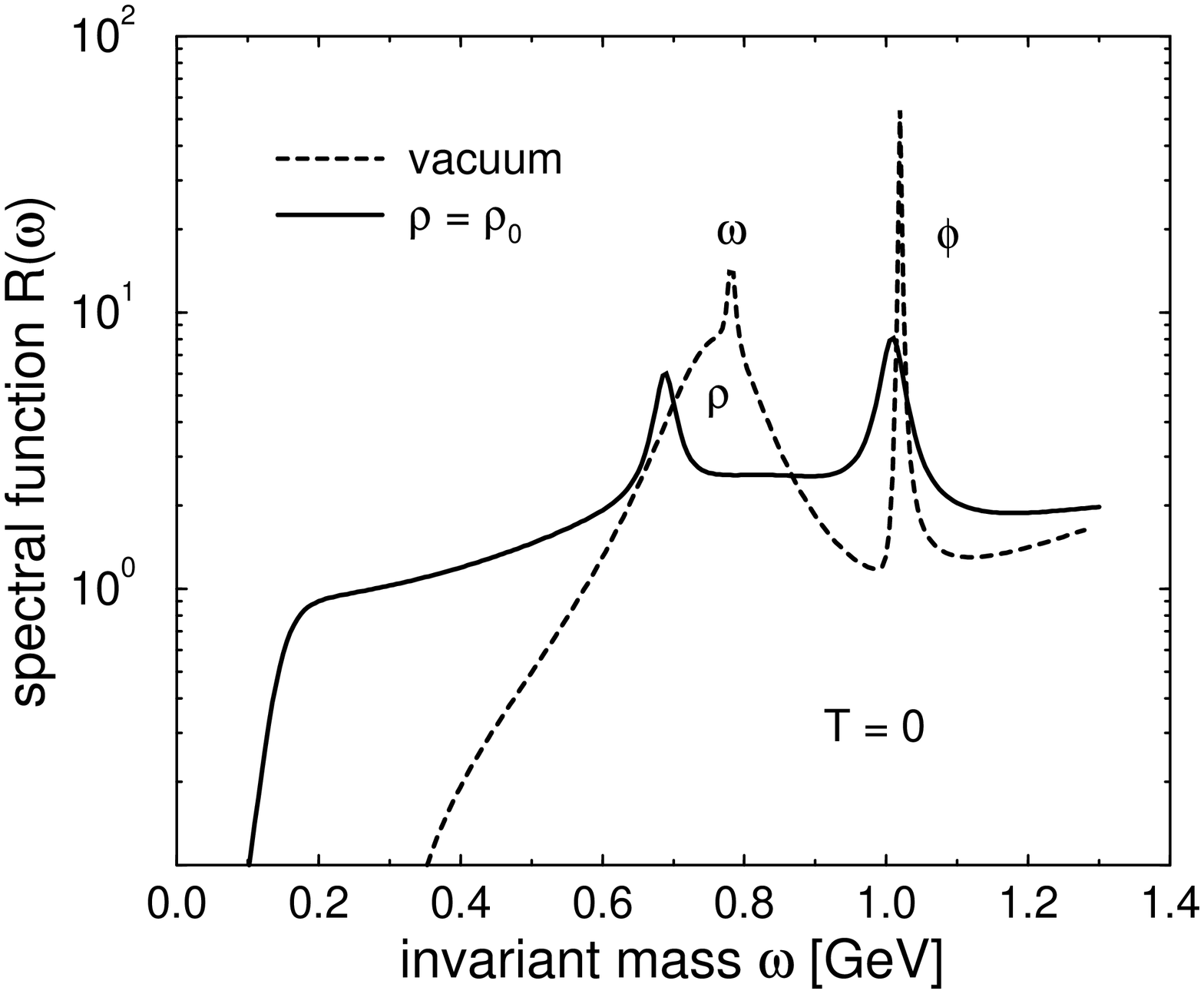,width=6.0cm, angle = -90}
\end{center}
\caption{The photon spectral function $R(\omega) = (12\pi/\omega^2) \ \mbox{Im}
 \bar{\Pi}(\omega)$ at finite temperature and $\rho_B = 0$ (left panel) and at
$T=0$ and baryon density  of normal nuclear matter, $\rho_B = \rho_0 = 0.17$ fm$^{-3}$
(right panel). For orientation, the $q\bar{q}$ line in the left panel
shows the spectral function in the QGP phase with  massless $u$- and $d$-quarks and
$m_s = 150$ MeV for $s$-quarks, neglecting $\alpha_s$-corrections.}
\label{hadron_spectra}
\end{figure}

Finite temperature modifications of the vector meson self-energies appearing
in eq.(\ref{ImBarPi}) are calculated using thermal Feynman rules. The explicit
calculations for the $\rho$- and $\phi$-meson can be found in ref.\cite{SW00}.
At the one-loop level, the $\rho$ and $\phi$ are only marginally affected by
temperature even close to $T_C$ because of the comparably large pion and kaon
masses: $m_\pi \simeq T_C$, $m_K \simeq 3 \ T_C$. The thermal spectral
function of the $\omega$-meson has been discussed in detail in \cite{SW01}.
Here, the reaction $\omega \pi \rightarrow \pi\pi$ was found to cause a
considerable broadening of the $\omega$ spectral function, leading to
$\Gamma_\omega(T_C) \simeq 7 \ \Gamma_\omega(0)$. The corresponding photon
spectral function is displayed in figure \ref{hadron_spectra} (left panel).

At higher invariant masses 1 GeV $ < M < $ 2 GeV, $\pi a_1$ 
annihilation is the dominant source of dileptons \cite{A1_1, A1_2}. The
vacuum vector and axialvector spectral functions become mixed to order
$T^2$ with a strength $T^2/(6 f_\pi^2)$ ($f_\pi \simeq 93$ MeV is the
pion decay constant) due to their coupling to the pionic heat bath
\cite{EJ} and should be degenerate at the point of chiral symmetry
restoration. The effect of the $a_1$ and higher resonances can then be
approximately subsumed in a structureless continuum above 1 GeV
\cite{MIXA1}.
We practically implement the $\pi a_1$ contribution by adding a flat
$2\pi$ continuum to the resonance $\rho$ meson spectral function that
feeds into the photon spectral function, reminiscent of the perturbative
plateau of $q\bar{q}$ annihilation.

There is still considerable stopping of the interpenetrating nuclei at SPS energies,
resulting in a net baryon
density $\rho_B$ in the central rapidity region. At RHIC, measurements
\cite{RHIC1, RHIC2} indicate that the proton over antiproton excess is small,
implying that the baryons are distributed over a larger rapidity interval.
Therefore, finite baryon density effects should not play such an important role at
RHIC kinematics.

For the evaluation of density effects which are relevant at SPS conditions,
we use the
results discussed in \cite{KKW2}. There it was shown that in the linear
density approximation, $\Pi_V$ is related to the vector meson - nucleon
scattering amplitude  $T_{VN}$:
\begin{equation}
\Pi_V(q^0, \vec{q} = 0; \rho) = \Pi_V^{vac} - \rho_B T_{VN}(q),\quad T_{VN}(q)
= - \frac{i}{3} \int d^4 x   \ e^{iqx} \langle N | \mathcal{T} j_\mu(x)
j^\mu(0) | N \rangle, \label{PiV}
\end{equation}
with $|N\rangle$ being zero-momentum free nucleon states. In the
following, we assume that the  temperature- and density dependences of $\Pi_V$
factorize, {\em i.e.} we replace $\Pi_V^{vac}$  in eq.(\ref{PiV})
by the
temperature-dependent $\Pi_V(T)$ and leave $T_{VN}$ unaffected. This amounts
to neglecting contributions from matrix elements such as $\langle \pi N |
\mathcal{T} j_\mu(x) j^\mu(0) | \pi  N \rangle$ (nucleon-pion
scatterings where the pion comes from the heat bath). Furthermore, this
approximation does not take into consideration a possible $T$-dependent pion
or nucleon mass. Some effective models suggest that, near the
phase transition, the nucleon mass follows the behaviour of the chiral
condensate $\langle \bar{\psi} \psi \rangle$ and drops abruptly as the
quarks lose their constituent masses. Such modifications of particle
properties may have a considerable impact on the spectral functions. However,
since the temperature range over which the dropping takes place is narrow, we
expect such effects not to leave distinct signals in the dilepton spectra
which are only sensitive to the integrated time (and hence temperature)
evolution of the system.

The photon spectral function at finite density and
zero temperature is depicted in figure \ref{hadron_spectra} (right panel).
The interaction with nucleons causes a strong broadening of the $\rho$ meson down
to the one pion threshold, leading to a complete dissolution of its
quasiparticle peak structure. The modifications of the $\omega$ and $\phi$
meson spectral distributions are more moderate: The mass of the $\omega$
drops by about 100 MeV at normal nuclear matter density, and its width increases by a
factor of about 5, whereas the $\phi$ mass stays close to its vacuum value,
accompanied by a ninefold increased width.

To summarize, the most prominent changes of the photon spectral function, when
compared to the vacuum case, arise from the broadening of the $\rho$ due to
finite baryon density effects and the broadening of the $\omega$ due to
scattering off thermal pions. The $\phi$ meson retains its distinct
peak structure even under extreme
conditions of density and temperature.
Very close to $T_C$, however, these results based on perturbative
calculations, are not expected to be reliable.

\subsection{After freeze-out contributions}
\label{subsec_after}
At the freeze-out stage, there are still vector mesons present. These will 
decay with their vacuum properties on their way to the detector and add to 
the dilepton yield from the previous thermalized phase. The invariant mass 
region below approximately 400 MeV is mainly filled by the Dalitz decays 
of the vector mesons. We take these contributions from the experimental 
analysis of the CERES collaboration  for SPS conditions. Since the 
PHENIX acceptance starts only above 1 GeV, the Dalitz decays do not 
play a significant r\^ole at RHIC.

For the calculation of the  direct decay of a vector meson $V$ into a lepton
pair we start with the following formula:
$$
\frac{dN_V}{dM d\eta} = \frac{1}{\Delta \eta} \ \frac{\alpha^2}{12\pi^4} \ R_V(M, T=0) 
\int_{\tau_{f}}^\infty d\tau \ V_{f} \int d^3q \ \frac{M}{q^0} \ f_B(q^0, T_{f}) 
\exp\left(- \frac{\tau - \tau_{f}}{\gamma(q)\tau_0^V}\right).
$$
Here, $T_{f}$ and $V_{f}$ are the fireball temperature and volume, respectively, at freeze-out. 
After $\tau_{f}$, the freeze-out time, all medium effects are switched off, so the vacuum 
photon spectral function $R_V(M, T=0)$ determines the rate. The corresponding momentum 
distribution is given by the thermal Bose function, evaluated at the freeze-out 
temperature $T_{f}$. However, the {\em absolute number} $N$ of vector mesons as a 
function of time is not a constant: since the mesons decay and there is no thermal 
recombination, $N$ decreases exponentially like 
$\exp\left(- (\tau - \tau_{f})/(\gamma(q)\tau_0^V)\right)$. The vacuum life time of the 
vector meson $V$ under consideration is denoted by $\tau_0^V$, and $\gamma(q)$ 
accounts for time dilatation effects on particles with finite three-momentum:
$$
\gamma(q) = \frac{1}{\sqrt{1 - v^2}} = \frac{q^0}{M}.
$$
After the time integration we end up with:
$$
\frac{dN}{dM d\eta} = \frac{1}{\Delta \eta} \frac{\alpha^2}{12\pi^4}  \tau_0 R(M, T=0)  V_{f} \int d^3q \ f_B(q^0, T_{f}).
$$
The averaged space-time four-volume that is available after freeze-out is therefore $V_{f} \tau_0^V$, 
as anticipated. The integral over $d^3q$ now yields the freeze-out 
particle density $n(M) = N(M)/V$ of the virtual photons as a function of invariant mass. 
Note that the information  on the vector mesons remains {\em entirely} in the photon spectral function. 
With the factor $V_{f}$ we obtain the total number of photons at freeze-out. When weighted with $R_V(M)$, this gives the $dN/dM$ distribution for the process meson $\rightarrow 
\gamma^* \rightarrow e^+e^-$.

We have also checked that a commonly used Breit-Wigner {\em ansatz} of the form
\begin{equation}
\frac{dN_{e^+ e^-}}{dM} = \left[ V_{f}  n(T_f,M) \right]  \xi \ B_V \frac{M^2 \
\Gamma_V(M)}{(M^2 - m_V^2)^2 + M^2 \Gamma_V(M)^2}\label{final2}
\end{equation}
yields almost identical results. Here, $\Gamma_V(M)$ stands for the $M$-dependent decay width, $\xi$ is the normalization factor and $B_V$ the branching ratio for the leptonic decay of the corresponding vector meson.

\subsection{Drell-Yan and charm contributions}
\label{subsec_dy}

At invariant masses $M>1$~GeV the Drell-Yan mechanism, i.e., hard quark-antiquark annihilation $q\bar{q}\to\ell^+\ell^-$ at leading order (LO), constitutes another source of (non-thermal) dileptons. Its differential LO cross section in a nucleus($A_1$)-nucleus($A_2$) collision reads
\begin{eqnarray}\label{dy_cross}
& & \frac{d\sigma (A_1 A_2)}{dy\, dM} =  \frac{8 \pi \alpha^2}{9 \,M\, s} \sum_q e_q^2\,\times \\ 
&& \biggr[ \left(Z_1 \, f_q^p(x_1) + (A_1-Z_1)\,f_{q}^n(x_1)\right) \left(Z_2 \, f_{\bar{q}}^p(x_2)
+ (A_2-Z_2)\,f_{\bar{q}}^n(x_2)\right) + (q \leftrightarrow \bar{q}) \biggr] \nonumber
\end{eqnarray}
where $\sqrt{s}$ denotes the c.m. energy of the nucleon-nucleon collision and $x_{1,2} =  M/\sqrt{s}\, \exp(\pm y)$ is the momentum fraction of the beam and target parton respectively. 

The Drell-Yan cross section~(\ref{dy_cross}) is computed using the LO MRST parameterization~\cite{mrst} for the parton distributions $f_i^p(x, \mu^2)$ evaluated at the hard scale $\mu^2 = M^2$. However, it has been checked that using different LO sets (e.g., CTEQ5L~\cite{cteq} or GRV98LO~\cite{grv}) affects the results by only 10 \% at SPS and 20\% at RHIC energies. To account for higher order corrections, we multiply the LO expression (\ref{dy_cross}) by a $K$ factor $K = 2$ fitted from $p$--$p$ data~\cite{e772}. Finally, nuclear effects like shadowing or quark energy loss are expected to suppress the Drell-Yan yield by about 30-50\% \cite{EKS98, energy_loss}. Since these effects are still poorly known quantitatively, we neglect them and consider our Drell-Yan rate as an {\em upper limit} on the actual rate.

Using the Glauber model of multiple independent collisions, the average dilepton multiplicity in a $A_1$-$A_2$ collision at impact parameter $b$ is given by
\begin{equation}\label{DY_mult}
\frac{dN(A_1 A_2)}{dy\, dM} (b) = T_{A_1 A_2}(b)\times \,K\,\frac{d\sigma (A_1 A_2)}{dy\, dM},
\end{equation}
where the normalized thickness function $T_{A_1 A_2}(b)$ is computed assuming the standard Woods-Saxon nuclear density profile. The Drell-Yan pair multiplicity (\ref{DY_mult}) is then averaged for the 30\% and 6\% most central collisions to be compared with CERES and PHENIX data, respectively\footnote{In addition to these centrality cuts, we need to rely on further assumptions to take properly into account the acceptance of these experiments. Therefore, we shall assume the generic form $d\sigma/dp_\perp \propto p_\perp / (1+(p_\perp/p_0)^2)^6$ ($p_0 = 3$~GeV) for the $p_\perp$ dependence of the DY process~\cite{e772}. Furthermore, neglecting corrections due to the intrinsic $k_\perp$ of the incoming partons, the angular distribution is taken to be $dN/d\Omega \propto 1+\cos^2\theta$ where $\theta$ is the angle between the lepton and the beam axis.}.

Another source of dileptons in the high invariant mass region consists of semileptonic decays of charmed mesons. Whereas earlier calculations found a considerable yield from open charm exceeding the thermal radiation \cite{OC0}, the subsequent inclusion of medium effects like energy loss led to a suppression of the dielectron rate and made it comparable to or even lower than the Drell-Yan yield \cite{OC1, OC2}. Since the Drell-Yan contribution plays only a subdominant r\^ole in the following, we will not explicitly include the open charm contributions.

\section{Fireball model}
\label{sec_fireball}

\subsection{General properties}

We do not aim at the detailed description of the
heavy-ion collision or its subsequent expansion on
an event by event basis or by a hydrodynamical simulation.
Instead we use a model of an expanding fireball which
enables us to test different scenarios with
different parameter sets in a systematic way, so as
to gain insight into
the time evolution of the strongly interacting system.

We assume that the physics of the fireball is the same
inside each volume of given proper time $\tau$, thus
averaging over spacial inhomogenities in density and
temperature. The volume itself is taken to
be an expanding cylinder, in which the volume elements
move away from the center in order to generate the
observed flow. There is no global Lorentz frame in which
thermodynamics can be applied. As the fireball
expands, volume elements away from the center are
moving with large velocities and are subject to time
dilatation when seen in the center of mass frame
of the collision. We assume a
linear rise in rapidity when going from central
volume elements to the fireball edge along the beam ($z$)-axis and the transverse axis. As the velocities along the
$z$-axis are typically large (up to $c$) as compared to transverse motion
(up to 0.55 $c$) for SPS and RHIC conditions, we make the
simplifying assumption that the proper time is in a
one-to-one correspondence to the $z$-position of a
given volume element, thus neglecting the time dilatation
caused by transverse motion. The whole system is assumed to
be in local thermal (though not necessarily in chemical)
equilibrium at all times.

Given this overall framework, the volume expansion of the fireball
is governed by the longitudinal growth speed $v_z$ and the
transverse expansion speed  $v_\perp$ at a given proper time.
These quantities can be determined at the freeze-out point and
correspond to the observed amount of flow. However, flow
is measured in the lab frame and needs to be translated into
the growth of proper time volume.
We use a detailed analysis of the freeze-out conditions for
central Pb-Pb collisions at 160 AGeV \cite{FREEZE-OUT} to fix the endpoint of the
evolution. The initial state is constrained using the overlap
geometry of the colliding nuclei. The expansion between 
initial and freeze-out stages is
then required to be in accordance with the EoS as determined
from the quasiparticle model described in section \ref{sec_quasiparticles}.
The resulting model serves as the basic setup, its extension to
different beam energies and collision centralities is discussed in
\ref{sec-scale}.

\subsection{Initial and Freeze-out conditions}

We use the data set {\bf b1} of \cite{FREEZE-OUT} as the endpoint
of our fireball evolution. The data set has been obtained by
a simultaneous fit of the fireball emission function to
hadronic $m_T$ spectra and HBT radii, so as to
disentangle the contributions from flow and temperature to
these quantities.

The fireball is characterized by a transverse box-shaped density
distribution with a radius $R_B= 12.1$ fm. This corresponds to
a root mean square radius of $R_{rms}^f = 8.55$ fm ($=R_B/\sqrt{2}$). The average
transverse expansion velocity is found to be 0.5 $c$ and the
temperature at freeze-out is $T_f= 100$ MeV.

Looking at the longitudinal expansion, a velocity $v_z^f \approx 0.9 $ $c$
at the fireball front is needed in $\pm z$ direction in order to account for the
observed shape of $dN_{ch}/dy$ distributions.

The initial conditions in transverse direction can be fixed
by the overlap geometry. Here, $R_0\approx 4.5$ fm
and $v_\perp = 0$ for central collisions. The initial longitudinal size of the
fireball is related to the amount of stopping of the
matter in the collision and the time $\tau_0$ necessary
for the formation of a thermalized system. There is no
direct information on these two quantities. However,
pQCD calculations indicate $\tau_0 \approx 1$ fm/$c$ for
SPS conditions and shorter times at RHIC.
The initial longitudinal velocity
$v_z^i$ can be inferred from hydrodynamical calculations.
Since the initial state in principle determines the
final state (if the EoS is known), one can fit this parameter to
the observed $dN/dy$ spectra. This procedure 
points at $v_z^i \approx 0.5 \ c$.
The initial longitudinal size of the system
at proper time $\tau_0$ is then calculated
by the intercept of $z(t) = v_z^i \cdot t$ with the $\tau = \sqrt{t^2-z^2}= \tau_0$
line.

\subsection{Volume evolution}

Using the available information on initial and freeze-out
conditions and the EoS of the system, we are now able to reconstruct
the evolution of the fireball volume in proper time:\\[0.5cm]
(1) The EoS translates into a temperature (and hence $\tau$) dependent
acceleration profile $a = const. \cdot p(T) / \epsilon(T)$ that exhibits a
soft transition point at $T=T_C$.
This can be understood as follows: the thermal vacuum energy $B(T)$
adds to the energy density $\epsilon$ but is subtracted from the pressure $p$.
As evident from figure \ref{2_1_flav},  $B(T)$ gets large near the phase transition
and correspondingly the ratio $p/\epsilon$ drops. In the interpretation
of $B(T)$ as a 'bag pressure', the pressure of the
QGP becomes partly counterbalanced by the confining bag pressure. Comparing initial and final
conditions, we need both longitudinal and transverse acceleration to account
for the velocity gain, and we keep the possibility of having two different constants
$c_\perp$ and $c_z$ which relate transverse and longitudinal accelerations to $p/\epsilon$. In practice, the temperature evolution with
$\tau$ is calculated starting with a trial ansatz and iterating
the result to obtain a self-consistent solution.\\[0.5cm]
(2) Starting with an ansatz for the radial expansion velocity,
\begin{equation}
v_\perp(\tau) = \int_0^\tau d \tau' c_\perp\frac{p(\tau')}{\epsilon(\tau')}
\end{equation}
and
\begin{equation}
R(\tau) = R_0 + \int_0^\tau \int_0^{\tau'} d \tau' d\tau'' c_\perp\frac{p(\tau'')}{\epsilon(\tau'')}
\end{equation}
where $R_0$ is the initial overlap rms radius of the collision region, we
fix the two unknown parameters $c_\perp$ and $\tau_f$ by the requirement that
$R(\tau_f)=R_{rms}^f$ and  $v_\perp(\tau_f) = v_\perp^f$. \\[0.5cm]
(3) For the longitudinal expansion we follow the motion of the fireball front
in the center of mass frame and use the expressions
\begin{equation}
v_z(t) = v_z^i + \int_0^t d t' c_z\frac{p(t')}{\epsilon(t')}
\end{equation}
and
\begin{equation}
z(t)= z_0 + v_z^i \cdot t + \int_0^t \int_0^{t'} d t' dt'' c_z\frac{p(t'')}{\epsilon(t'')}.
\end{equation}
The parameters $c_z$ and $t_f$ (freeze-out in c.m. frame) are fixed by
$v_z(t_f) = v_z^f$ and $z(t_f)$ to lie on the $\tau = \tau_f$ line, with
$\tau_f$ determined from the radial expansion.\\[0.5cm]
(4) The movement of the fireball front in the c.m. frame can now be translated
into the growth of volumes in proper time by finding the intercept of
$z(t)$ and $\tau=const.$ and calculating the pathlength along this curve of
fixed proper time.

In order to construct a self consistent evolution model, 
we need to find the proper ratio $p(T(\tau))/\epsilon(T(\tau))$ corresponding to
the volume expansion.
The evolution of $T(\tau)$ is dealt within the next section. 
For the QGP phase, the ratio is then determined within the quasiparticle model.
Unfortunately, no reliable information on $p/\epsilon$ is available for the hadronic
phase. We can, however, deduce the value of this ratio at both
$T_C$ and $T_{f}$, and interpolate between these limits.
As the lattice indicates neither a sharp drop in $\epsilon$ nor in
$p$ when approaching $T_C$ from above, it appears reasonable to assume that
the $p/\epsilon$ ratio stays small in the vicinity of the phase transition
even in the hadronic phase.
On the other hand, the decoupling of the system at
freeze-out implies that interactions between its constituents
become unimportant, therefore we recover the ideal gas limit
at $T \rightarrow T_f$, where
standard thermodynamics predicts $p/\epsilon = 1/3$ (for massless particles).
We now interpolate linearly between these two values to
cover the temperature region inbetween. 

\subsection{Temperature evolution}

The temperature profile $T(t)$ of the fireball is uniquely determined
by  the assumed condition of isentropic expansion once the volume expansion is
known.
In order to derive the temperature $T$ at a given time,
we calculate the entropy density via
\begin{equation}
s(\tau) = S_0/V(\tau),
\end{equation}
where $S_0$ is the total entropy of the fireball. The relation between the entropy density $s$ and the
fireball temperature $T$ is fixed by the EoS. By inverting this (unique) relation numerically, 
we finally obtain the temperature profile.

We infer the EoS of the QGP phase from the thermal quasiparticle model \cite{RW01}
adapted to lattice QCD results.
Unfortunately, determining the behaviour of the hadronic 
phase is not quite as easy for a number of reasons. First,
lattice results  usually employ quark masses
which are far too large, leading to pion masses $m_\pi^{latt}\gsim 3\, m_\pi^{phys}$.
The resulting thermal suppression of these degrees of freedom
causes a considerable discrepancy of the lattice EoS with respect to that of
the interacting pion gas or the free hadronic resonance gas.
On the other hand, the description in terms of a non-interacting
or perturbatively interacting system is bound to fail near the
phase transition.
Only in the very final stages, near the freeze-out,
interactions cease to be important and one may assume that a
non-interacting system describes the situation adequately.

We parametrize our insufficient knowledge
close to $T_C$  by interpolating smoothly between two regimes, the
QGP quasiparticle result for $T>T_C$ and the noninteracting hadronic
resonance gas result for $T<T_{f}$. This approach is supported by the general
shape of the EoS on the lattice for two light quarks and one heavy
quark, which resembles a weak first-order or second order phase transition, or even
a smooth crossover \cite{KAR}.

\begin{figure}[htb]
\begin{center}
\epsfig{file=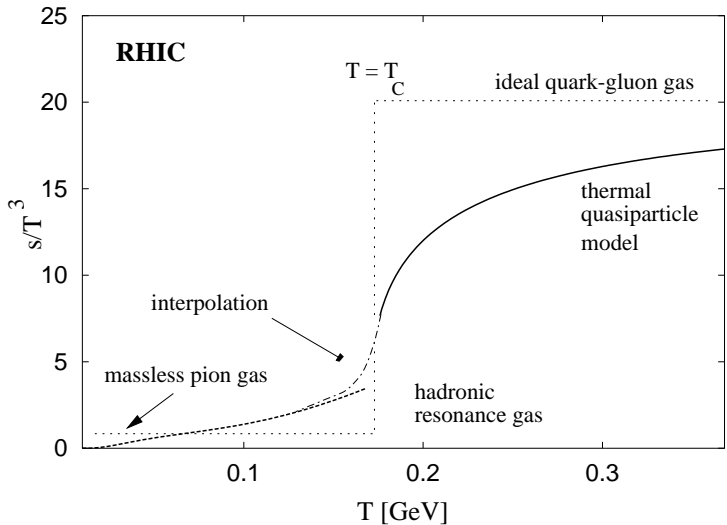, width=10cm}
\end{center}
\caption{\label{F-EntropyRHIC}The temperature dependence of the entropy density $s(T)$
in the RHIC kinematical scenario as compared to the ideal quark-gluon gas and the massless noninteracting
pion gas limit (dotted line). The three relevant regions used in the model
calculation are given as
ideal hadronic resonance gas (solid line), interpolation (dash-dotted line) and thermal quasiparticle
model (dashed line).}
\end{figure}

The situation is shown for the RHIC scenario in figure~\ref{F-EntropyRHIC} and compared
to the ideal quark-gluon gas and the hadronic resonance gas approach used in other works
({\em e.g.} \cite{HYDRO-SOLLFRANK, HYDRO-SHU}).  One can clearly
observe that the deviations in both the quark-gluon phase and the hadronic phase from the ideal gas
are large, amounting to a factor of more than two for the entropy density near $T_C$.
Keeping in mind
that a smaller entropy density translates into a higher temperature in an isentropic expansion,
we conclude that our model predicts a prolonged lifetime of the QGP phase as compared to
the ideal gas ansatz, whereas the lifetime of the hadronic phase is reduced somewhat.

At SPS, the measured ratio of protons over antiprotons indicates
a partial stopping of nucleons
during the collision phase, resulting in an excess of quarks over antiquarks
in the fireball region \cite{P-OVER-PBAR}. Since baryon number is conserved, this implies the
existence of a $T$-dependent baryochemical potential, which in turn
translates into a fugacity factor $\lambda_B = \exp[\frac{\mu_B}{T}]$
multiplying the entropy density created by baryons.
Furthermore, the thermal yield of pions in the fireball is not enough to
account for the observed number of pions. This can be compensated by the
introduction of a chemical potential for pions (and kaons), which in turn
influences the EoS.

By imposing entropy and baryon number conservation, the evolution of $\mu_B$ can
be followed through the fireball expansion. In practice, the evaluation at each $T$ rests on the
assumption of the system being an ideal hadronic gas, which we believe is unreasonable near the
phase transition. Fortunately, $T_f \simeq 100$ MeV is far enough distant from the phase transition,
and that is where we fix the entropy.
Furthermore, we expect pions to be the dominant thermally active degrees of freedom.
We use a pion chemical potential 
$\mu_\pi(T_f) = 123$ MeV as obtained in \cite{FREEZE-OUT}, which gives the
correct total pion multiplicity when evaluated with the fireball freeze-out volume.

\begin{figure}[htb]
\begin{center}
\epsfig{file=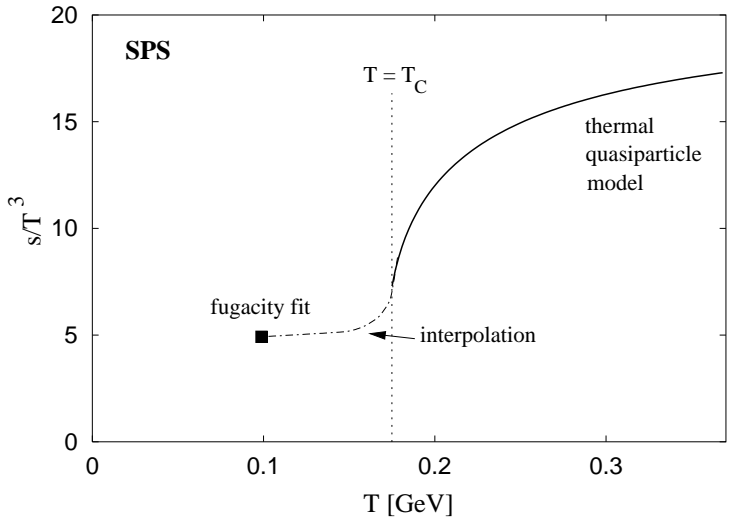, width=10cm}
\end{center}
\caption{\label{F-EntropySPS}. Temperature dependence of the entropy density for
SPS conditions, including the interpolation to the fugacity corrected value
at freeze-out. As the entropy density below $T_{f}$ is irrelevant for
the fireball evolution, the interpolation is stopped at this point.}
\end{figure}

Correcting the contributions from the different particle species for the corresponding
fugacity factors, we obtain
a point of the $s(T)$ curve where our interpolation starts.
This situation is shown in figure~\ref{F-EntropySPS}.
One observes that, unlike the RHIC scenario, the entropy density
under SPS conditions is larger
in the range $T_f<T<T_C$,
resulting in a faster dropping of temperature during the hadronic expansion phase.

Having now specified the relevant part of the EoS of the system, the temperature
evolution of the fireball is uniquely determined by the total entropy $S_0$
and the evolution of the volume $V(\tau)$ with proper time. The total entropy can be
 obtained by measuring charged particle multiplicities $N^+$ and $N^-$ 
in suitable rapidity bins and
calculating
\begin{equation}
D_Q = \frac{N^+ - N^-}{N^+ + N^-} \label{D_Q}.
\end{equation}
The quantity $D_Q$ stands for the inverse of the specific entropy 
per net baryon $S/B$, and the product $D_Q(S/B)$ roughly measures the entropy 
per pion \cite{ENTROPY-BARYON}. For SPS collisions at 160 AGeV, we find $S/B = 26$ for 
central collisons. For RHIC 6\% central Au-Au collisions at 130 AGeV, the specific 
entropy $S/B = 130$ is substantially higher due to the larger particle multiplicity 
and the smaller net baryon content in the central region. For beam energies of 
200 AGeV that are of interest here, not enough information on $dN/dy$ spectra 
is available at this moment, so we utilize the predictions from a thermal model 
calculation \cite{RATIOS-RHIC}. With the ratios 
$\bar{p}/p = 0.75$, $\bar{p}/\pi^- = 0.09$ and $K^-/\pi^- = 0.15$, we obtain 
the specific entropy as $S/B \approx 250$, as already estimated (albeit with 
different parameters) in \cite{RHIC}.

\subsection{Variations with beam energies and centralities}

\label{sec-scale}

It is mainly the detailed information on the final and initial
state which enables us to reconstruct the fireball evolution
in the case of the 160 AGeV central collision scenario.
Unfortunately, no such detailed analysis is available for the
30\% central PbAu collision scenario for which dilepton data
have been taken. The same holds for the 40 AGeV dilepton data.
It is therefore necessary to extend the
framework established so far to different centralities and
beam energies by a suitable re-scaling of key quantities characteristic
of the evolution.

First of all, the total entropy deposition in the fireball region
must be different when going to more peripheral collisions
or lower beam energies. We assume that the entropy per baryon scales
with the number of negatively charged hadrons observed in the final state and take the SPS value of $s/\rho_B = 26$ as a baseline.
When going to peripheral collisions, we assume that this
value is still a good estimate. Here, the total entropy in the system
is naturally reduced because the number of nucleons
 participating in the collision is smaller.

Second, the overlap geometry is different in peripheral collisions, resulting
in a smaller initial fireball volume. Here, we neglect details of the
transverse geometry and keep parametrizing the fireball volume as a cylinder.
Its initial transverse area is adjusted to the value of the calculated overlap
area, hence the cylinder radius is reduced as compared to central
collisions.  

For the 40 AGeV data, we take the total entropy to be about half the entropy at
160 AGeV because $N_{\pi^-}^{160} \approx 600$ and $N_{\pi^-}^{40} \approx 310$.
The initial energy density, estimated by Bjorken's formula from $dN/dy$ at midrapidity,
is about $2/3$ of the value at 160 AGeV, but still well above the critical energy
density required to form a QGP plasma. Looking at the final state of the fireball,
HBT analyses of NA49 \cite{40_HBT_NA49} indicate that the radius parameters are
very similar at 40 and 160 AGeV, suggesting that the reaction dynamics do not
significantly change in this energy region.

Changes in the initial entropy deposition result in a
correspondingly different fireball evolution and in general a different
freeze-out state.
Freeze-out occurs when the mean free path of particles exceeds the
dimensions of the fireball. As pions are the most abundant particle species
in the fireball, we assume that the pion density determines the mean free path 
$\lambda$ of particles in the medium. The freeze-out condition reads therefore
\begin{equation}
\label{E-FO}
\sigma \lambda \rho_\pi = 1,
\end{equation}
where $\sigma$ stands for a typical hadronic cross section. As the observed HBT radii for 40 and 160 AGeV appear rather similar,
we do not expect the total freeze-out volume to change more than
a factor two. This is thermodynamically consistent with fixing
the freeze-out temperature $T_f = 100$ MeV for all SPS scenarios. 
As $\lambda$ scales $\sim \sqrt[3]{V}$, this is a sensible ansatz --- the resulting
freeze-out pion densities are very similar when looking at
(\ref{E-FO}). This is still true if we take modificatios of the pion density
by a pion chemical potential $\mu_\pi$ into account --- 
about the same value of $\mu_\pi$ is needed
in all SPS scenarios to account for the observed total pion yield (see section
\ref{Chemical}).

The situation is different at RHIC. Here, no large pion chemical
potential appears to be necessary to account for the observed
total yield and therefore the thermal pion density
at a given temperature is lower, leading to a higher
freeze-out temperature with (\ref{E-FO}).
We find that an expansion scenario with $T_f$ =
130 MeV fits both the observed total particle yield
and leads to the correct freeze-out pion density.
The freeze-out volume has now to be adjusted
accordingly. In order to do this consistently, we simultaneously
modify both the final state flow velocities and the fireball
radius as compared to our standard scenario. 
As these quantities result to a first approximation
from an accelerated motion, reducing the radial flow
by a factor $f$ implies a reduction of the longitudinal flow
difference between initial and final state by the same factor $f$
and a reduction of the freeze-out radius by $f^2$.
Note that a combination of decreasing geometrical radius
\emph{and} flow velocity tends to keep the HBT radius
constant.

The fireball evolution is now calculated as described above,
using re-scaled total entropy, initial radius, final state longitudinal
and transverse flow and freeze-out radius as new inputs. The parameter
sets obtained for the different fireball scenarios are summarized in the
appendix.

\subsection{Results and discussion}

\begin{figure}[!htb]
\begin{center}
\epsfig{file=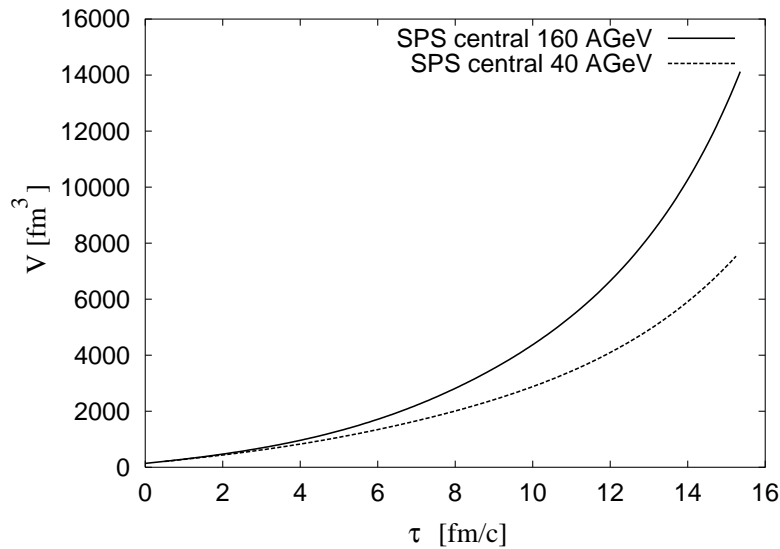, width=10.5cm}
\end{center}
\caption{\label{F-vc2} The volume expansion for SPS conditions. Shown
are the curves for central 160 AGeV collisions (solid line)
and central 40 AGeV collisions (dashed line).}
\end{figure}

\begin{figure}[!htb]
\begin{center}
\epsfig{file=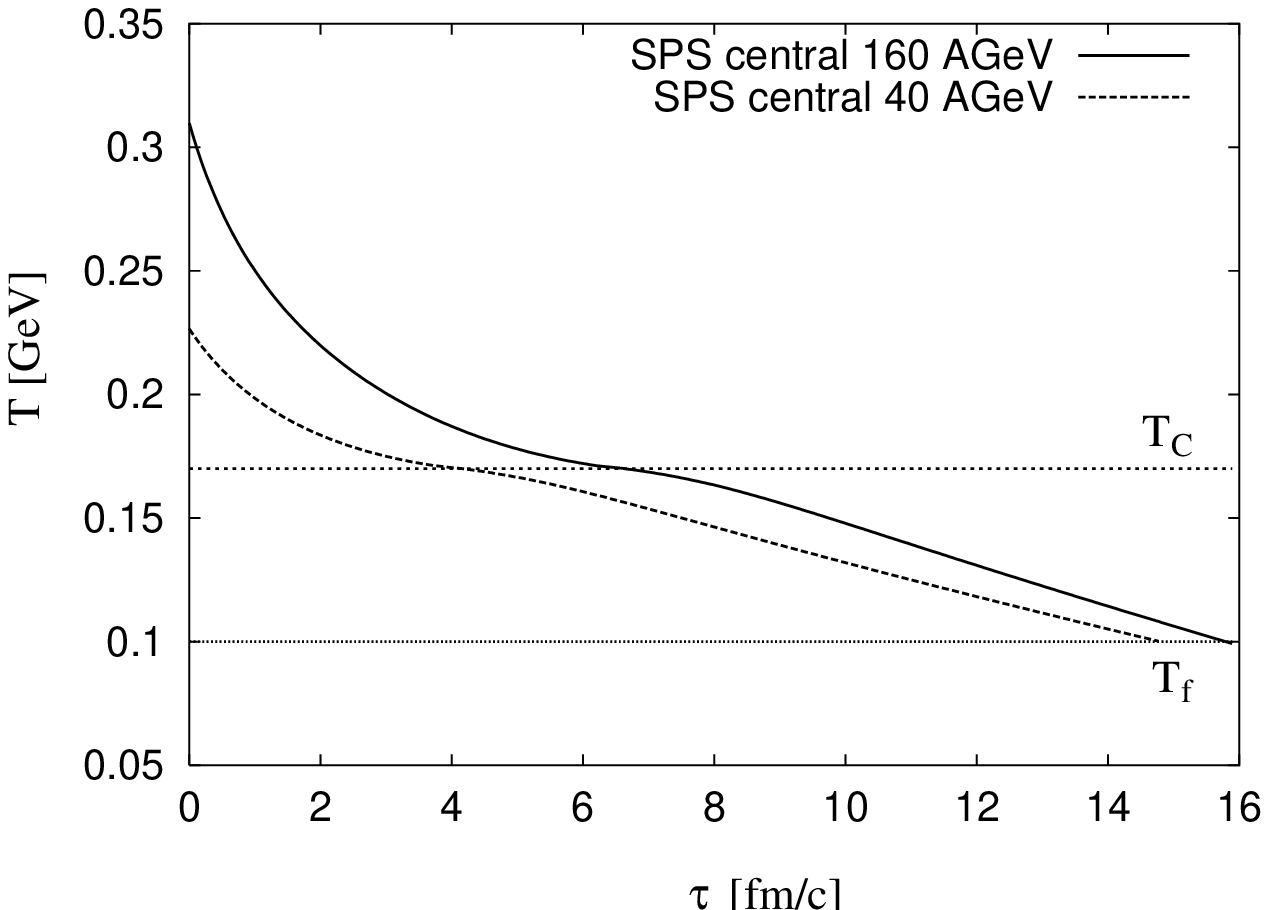, width=10cm}
\epsfig{file=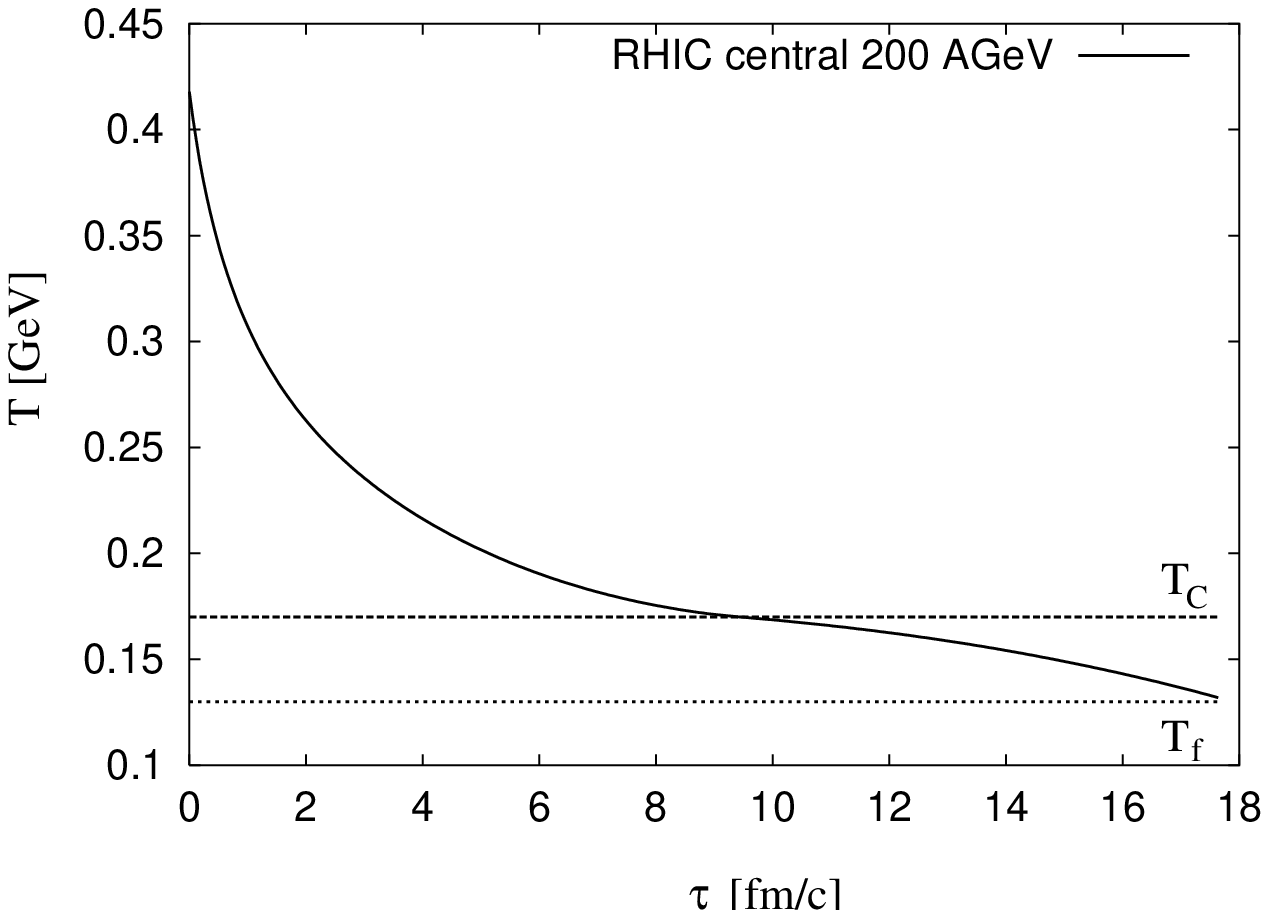, width=10cm}
\end{center}
\caption{\label{F-TProfiles} Time evolution of the temperature for SPS and
RHIC conditions as obtained with the fireball model described in this work. The critical and freeze-out temperatures are indicated by the horizontal lines.}
\end{figure}

The resulting volume evolutions for 40 and 160 AGeV are plotted 
in figure~\ref{F-vc2}.
Note that these curves correspond to the volume at proper time
$\tau$, which is larger than the geometrical size of the fireball
in the c.m. or the lab frame.
The temperature evolution is shown in figure~\ref{F-TProfiles}.
We observe that the QGP phase lasts about half of the total
fireball lifetime for 160 AGeV, and considerably less for 40 AGeV.
As the fireball evolution resulting from our model differs somewhat
from the results obtained by other groups, we feel that some
clarifying remarks are in order.

In our quasiparticle model there is no mixed phase of coexisting hadronic gas and QGP.
Based on the observation that there is \emph{no
strong first order} transition visible in the lattice data, the EoS
right and left of $T_C$ match smoothly. Inserting the EoS of an
ideal quark-gluon gas instead,
a large gap in the $s(T)$ diagram is created.
The resulting latent heat $(\Delta S) T_C$ generates a mixed phase
of considerable duration (5 - 10 fm), as found in previous approaches.

By construction of the model, hadronic observables derived from the
expanding fireball,
{\em e.g.} rapidity distributions of charged particles, ${\bf p}_t$-spectra, HBT radii
and observed particle numbers, are described adequately.
This is an important constraint of the model. There is very little freedom left to tune the fireball evolution. Arguably, the EoS
inferred from the quasiparticle model is the weakest point in the
chain of arguments leading to the complete picture. However,
bulk quantities such as freeze-out proper time and freeze-out volume
are hardly influenced by the EoS or the phase transition temperature. They can be found even by assuming constant acceleration at all times.

The initial temperatures are quite large ($\sim $ 300 MeV for SPS central 
conditions, 420 MeV for RHIC) and are uniquely determined by the total entropy, 
the initial volume and our EoS for the quark-gluon phase.
To some extent these large values related to our choice of the formation time 
$\tau_0 = 1.0$ fm/$c$ (SPS) and $\tau_0 = 0.6 $ fm/$c$ (RHIC) of the thermal
system; if $\tau_0$ is increased, the initial temperature decreases correspondingly.
However, unlike the results found in \cite{RHIC}, we do not observe any
strong sensitivity of the dilepton yields to the choice
of $\tau_0$: as the fireball expands,
differences in the initial volume become increasingly unimportant.
If the total evolution time of the QGP phase is small, these changes may well
matter, but as the lifetime of the QGP phase in our model is comparatively large,
the possible difference affects only a small fraction of the total lifetime
in a region where the fireball volume (and, correspondingly the dilepton
yield) is small anyway.  

Comparing with the Bjorken estimate of the initial energy density, one should keep
in mind that the extrapolation from the final to the initial state
is different in the present approach. In the Bjorken scenario, no
longitudinal acceleration is present, therefore the mapping of
final state rapidity distributions to initial state spatial distributions
finds a larger initial volume than our scenario. If we assume
no or only small longitudinal acceleration in order to
compare the two approaches, we find initial temperatures
between 220 and 240 MeV, consistent with the Bjorken estimate. 
As this assumption is incompatible with
a freeze-out at 100 MeV and the measured freeze-out geometry,
we disregard it.

We observe a  prolonged lifetime of the
QGP evolution phase of the fireball as compared to the results obtained by
other groups. Recall that the apparent lifetime of the QGP, as observed in the lab frame, is 
larger than its lifetime in proper time $\tau$. This is a consequence of our volume evolution 
and the use of the more realistic
EoS of our quasiparticle model, as opposed to that of the ideal quark-gluon gas.
Near the phase transition, the entropy density in the quasiparticle model
is about a factor two smaller than the one of the ideal quark-gluon
gas, and it takes
a correspondingly larger volume (and larger evolution time) to
reach this entropy density in an isentropic expansion.

The slower cooling time in the hadronic phase of the RHIC scenario as
compared to SPS conditions can in essence be traced back to the fact that
the entropy density under SPS conditions is enhanced by the fugacity factors
whereas this is not such an important effect for RHIC.

\subsection{Chemical composition}

\label{Chemical}

As already mentioned, the total thermal pion yield of the fireball
at freeze-out is not enough to account for the observed number of
pions, and a large chemical potential of $\mu_\pi = 123$ MeV has to
be used to compensate \cite{FREEZE-OUT}. The
common point of view is that particle numbers
are fixed at a chemical freeze-out point \cite{EQ2}
and the absence of inelastic
reactions introduces then a chemical potential. 
The statistical model is
enormously succesful in describing the finally observed
ratio as a fit of chemical freeze-out temperature and
baryon density, using the free properties of 
particles. However, in the hot and dense environment of
a fireball, in medium modifications of particle masses
and widths are likely to occur, changing the amount
of particle production at the freeze-out temperature.
This is not in direct contradiction to the apparent
success of the statistical model, as a different set
of $T$ and $\mu$ might be able to explain the observed
ratios once the modifiactions are taken into account.
Furthermore, the absence of all inelastic reactions,
leading to decay processes only, is most likely an
oversimplification for some particles, e.g. for the
$\rho \leftrightarrow \pi\pi$ system. If one knew
the particle properties and the particle abundancies
at hadronization reliably, the correct way to proceed
would be to set up a system of coupled rate equations
and follow the various decay and recombination processes
to the thermal freeze-out. Unfortunately, theoretically
there is no way to assess these properties reliably
near the phase transition, as this is a nonperturbative
problem. Keeping these issues in mind, we will therefore
parametrize the hadrochemistry of the fireball on a phenomenological
basis instead of aiming for a detailed solution of the
problem.

We adopt the following approach:
A pion chemical potential is introduced \emph{ad hoc} and fitted at the
thermal freeze-out point ($\mu_\pi = 123$ MeV \cite{FREEZE-OUT}) 
to the total pion multiplicity.
It is assumed to decrease with temperature linearly up to $T_C$ where
it vanishes, corresponding to
a situation where resonance decays continuously feed
pions into the system.
All other necessary chemical potentials are adjusted such as
to reproduce the observed pion to particle ratios at all times.

\section{Dilepton invariant mass spectra}
\label{sec_results}

Once the time evolution of the fireball is given in terms of the temperature $T(\tau)$, the baryon
density $\rho(\tau)$ and the volume $V(\tau)$, and with the knowledge of the photon spectral function,
we have all the necessary ingredients to calculate dilepton rates using eq.~(\ref{integratedrates}). We
fold the result with the acceptance of the CERES and the PHENIX detectors, respectively, and average
over the rapidity region covered by these two experiments. The so-called 'hadronic cocktail', dileptons
produced after freeze-out through various decay processes, with the exception of vector-meson decays,
is then added. This contribution fills the region of very low invariant masses ($M < 150$ MeV). The
dilepton yields resulting from direct vector meson decays after freeze-out, as described in subsection \ref{subsec_after}, and the Drell-Yan yield from subsection \ref{subsec_dy} are added to the hadronic cocktail, taking into account the limited kinematic acceptance and resolution of the detector.

\subsection{SPS data at 40 and 160 AGeV}
\label{subsec_sps}

\begin{figure}[!htb]
\begin{center}
\hspace*{-1.0cm} \epsfig{file=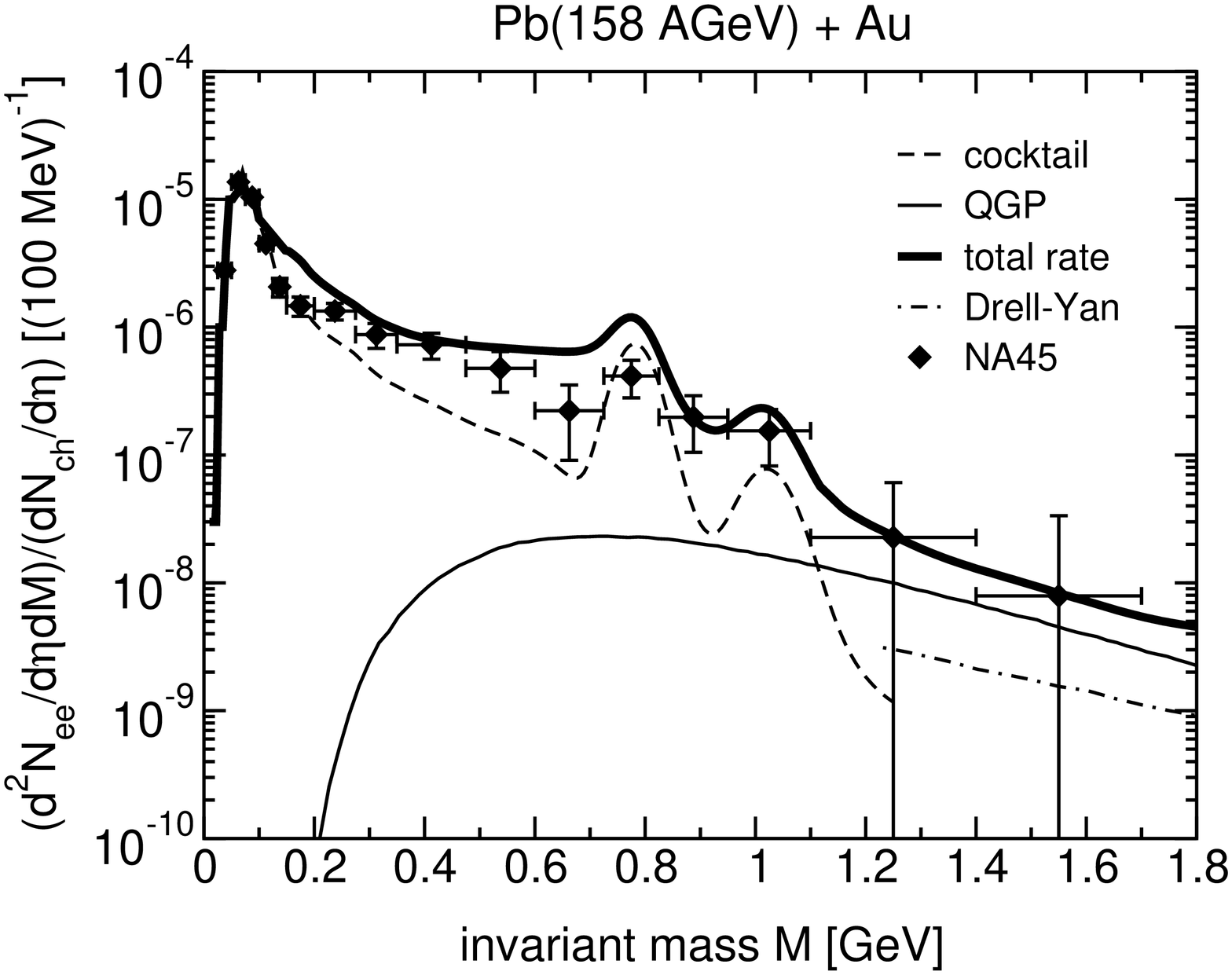, width=6.75cm, angle=-90} \hspace*{-1cm} \epsfig{file=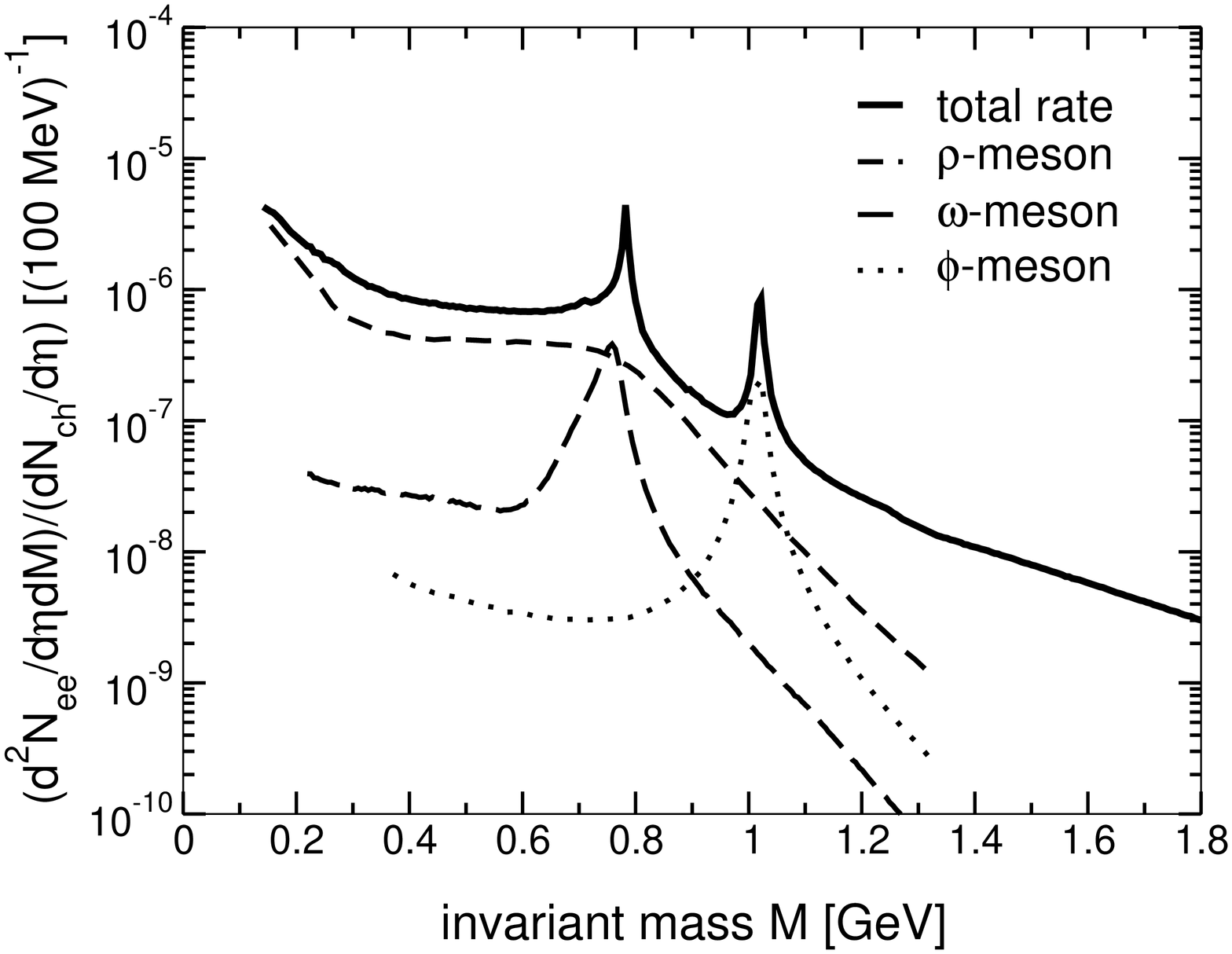, width=6.75cm, angle=-90} \hspace*{-2cm}
\end{center}
\caption{\label{Results_160}Left: Dilepton invariant mass spectra (see eq.~(\ref{integratedrates})),
normalized to $dN_{ch}/d\eta = 250 $, in units of (100 MeV)$^{-1}$, for the SPS CERES/NA45 Pb(158 AGeV)+Au 
experiment \cite{LENKEIT}. Shown are the data, the total rate, the cocktail
contribution including the after freeze-out decays of vector mesons, the QGP contribution and the Drell-Yan yield. Right: Contributions from $\rho$-, $\omega$- and $\phi$-mesons (excluding after freeze-out yield) shown separately, assuming perfect detector resolution.}
\end{figure}

\begin{figure}[!htb]
\begin{center}
\hspace*{-1cm} \epsfig{file=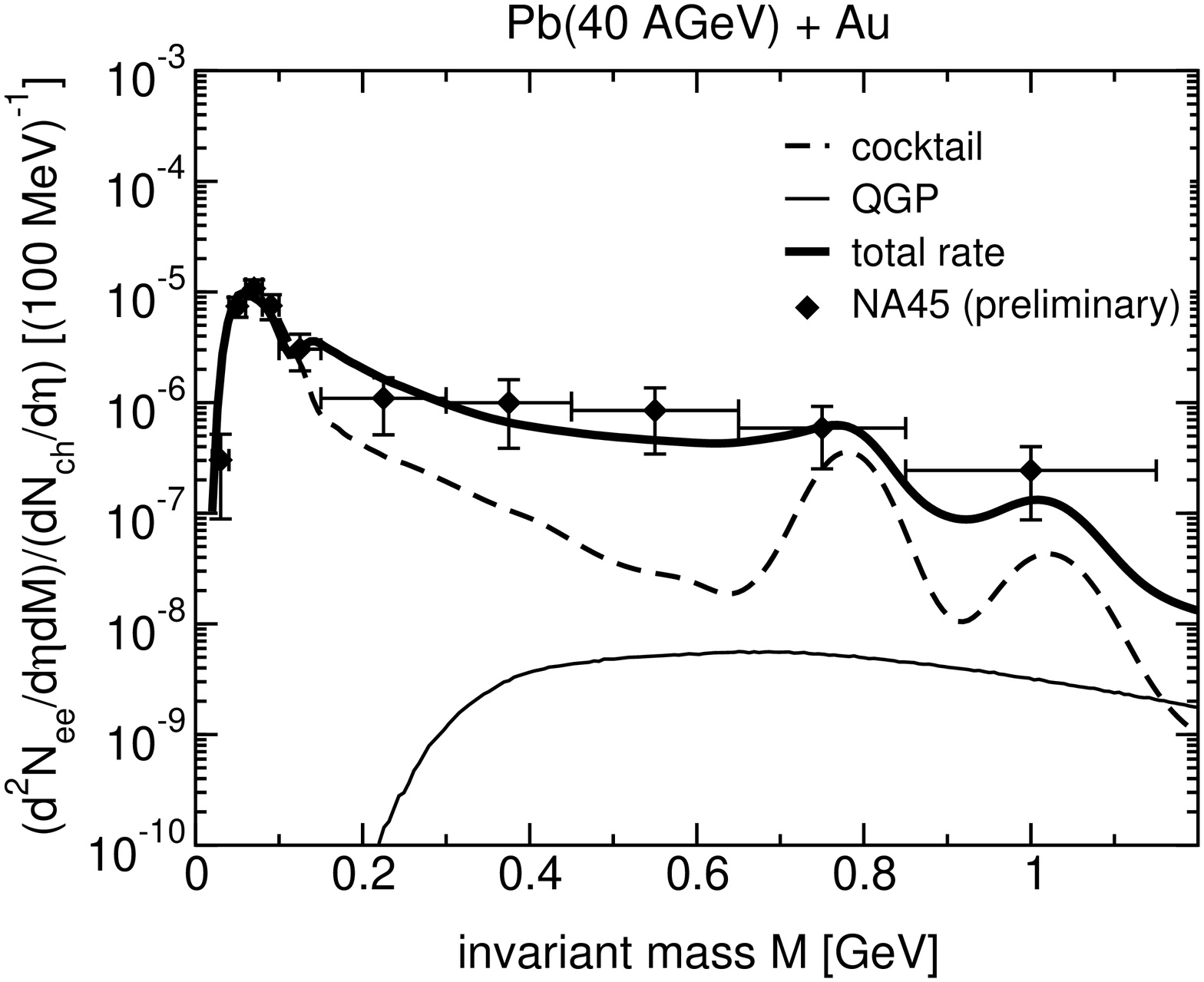, width=6.75cm, angle=-90} \hspace*{-1cm} \epsfig{file=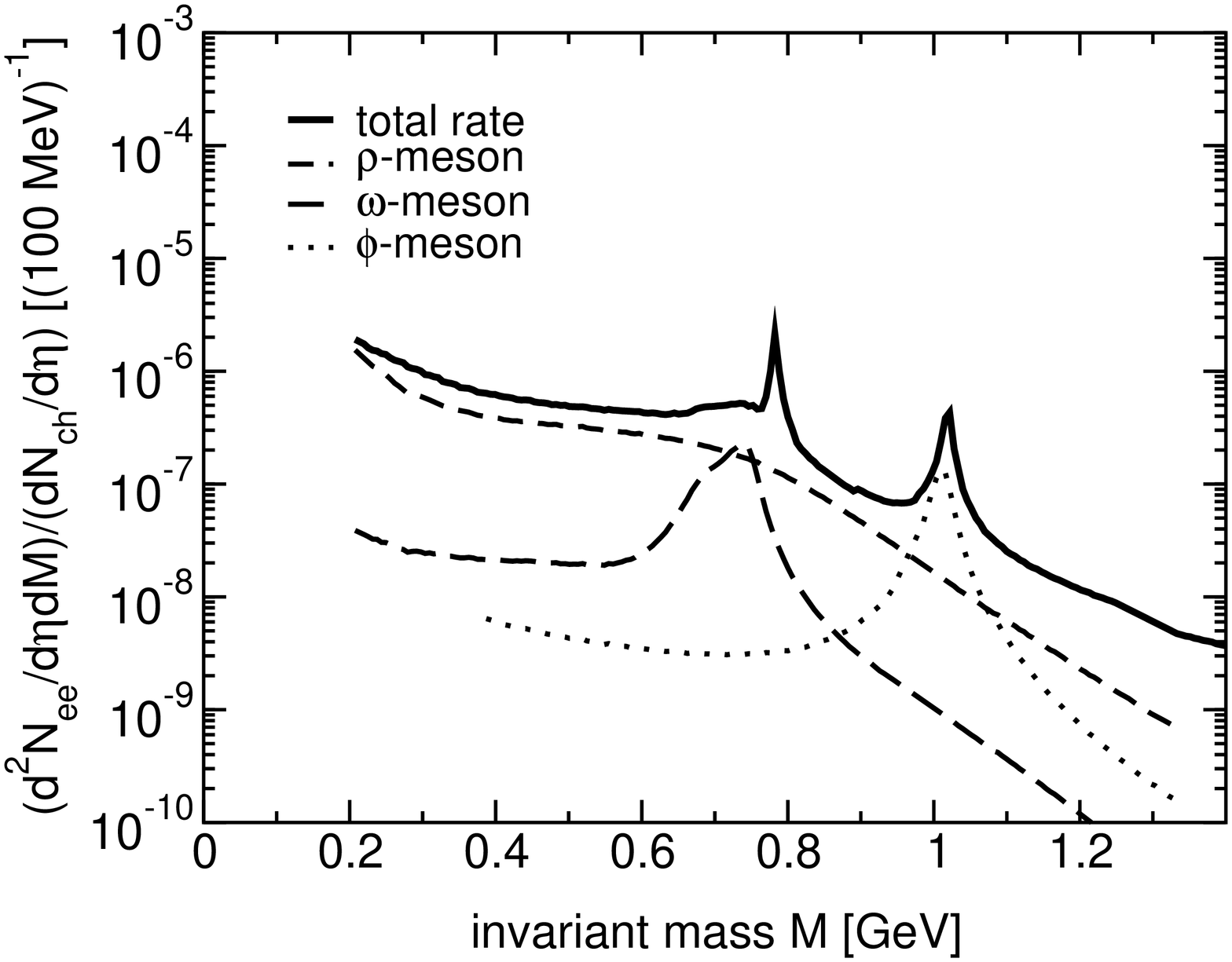, width=6.75cm, angle=-90} \hspace*{-2cm}
\end{center}
\caption{\label{Results_40}Left: Dilepton invariant mass spectra (see eq.~(\ref{integratedrates})),
normalized to $dN_{ch}/d\eta = 210 $, in units of (100 MeV)$^{-1}$, for the SPS CERES/NA45 Pb(40 AGeV)+Au experiment \cite{40GEV}. Shown are the preliminary data, the total rate, the cocktail contribution
including the after freeze-out decays of vector mesons and the QGP contribution.
Right: Contributions from $\rho$-, $\omega$- and $\phi$-mesons (excluding after
freeze-out yield) shown separately, assuming perfect detector resolution.}
\end{figure}

We start with a discussion of the SPS CERES/NA45 experiment, treating 40 AGeV and 160 AGeV data in
parallel. The final results for the dilepton invariant mass spectra are shown in
figs.~\ref{Results_160} and \ref{Results_40}.

Our calculation reproduces the overall spectrum of the 160 AGeV CERES data quite well. It overestimates
the rates somewhat around invariant masses of 200 to 400 MeV and achieves a good description in the
region above 400 MeV up to 1.8 GeV. Recall that our QGP model rate constitutes only a lower limit on the actual rate because it neglects
the radiation from non-partonic (cluster) degrees of freedom above the critical temperature. Bearing in
mind that the region above 1 GeV is mainly populated by dileptons originating from the QGP phase, as
evident from the left panel of figure \ref{Results_160}, there might still be additional radiation close above $T_C$ arising from hadronic clusters embedded in a QGP environment. The Drell-Yan contribution is non-negligible, but still outshined by the QGP by a factor of 3.

Changes in the spectra of the vector mesons indicate tendencies towards chiral symmetry restoration, so
the right panel of figure \ref{Results_160} shows the contributions of $\rho$, $\omega$ and
$\phi$ mesons separately, not including the after freeze-out yield. The $\rho$ meson loses its
quasiparticle structure entirely due to strong collision broadening at finite density, and fills the whole low
mass region between the two pion threshold and $\sim 800$ MeV. The $\omega$ meson, a sharp peak in the
vacuum, broadens at finite temperature due to the thermal scattering process $\omega \pi \rightarrow
\pi \pi$. Furthermore, the mass shift at finite baryon density smears the remaining peak structure
considerably. Effectively, the remaining signal arises from the direct decays of $\omega$ mesons
after freeze-out.  The $\phi$ has become a spread-out but still visible resonance structure, showing
only moderate broadening at finite temperature and baryon density. It might therefore be a suitable
candidate for gauging the strength of vector meson modification.

To test the modelling of the vector meson spectra, we calculate the total number of $\omega$ and $\phi$ meson, suitably averaged over their medium-induced spread in invariant mass, as
$$
\langle N_\omega \rangle = \frac{1}{N_{ch}} \int \limits_{0.65 \mbox{ \scriptsize GeV}}^{0.9 \mbox{ \scriptsize  GeV}} dM \ \frac{d^2 N_{\omega \rightarrow ee}}{d\eta dM} = 9 \cdot 10^{-7}
$$
and
$$
\langle N_\phi \rangle = \frac{1}{N_{ch}} \int \limits_{0.95 \mbox{ \scriptsize GeV}}^{1.1 \mbox{ \scriptsize GeV}} dM \ \frac{d^2 N_{\phi \rightarrow ee}}{d\eta dM} = 1.8 \cdot 10^{-7}.
$$
Comparing with numbers from a statistical model calculation, $\langle N_\omega \rangle = 4 \cdot 10^{-7}$ and $\langle N_\phi \rangle = 2.2 \cdot 10^{-7}$ \cite{SPECHT}, we indeed find reasonable agreement. Note that the relatively large $\omega$ meson yield is primarily caused by the pion fugacity factor $[\exp(\mu_\pi/T_f)]^3$ at freeze-out, which reflects the enhanced feeding through the $3\pi \rightarrow \omega$ process.

Going from 160 AGeV to 40 AGeV beam energy, analyses of HBT radii and transverse radial flow \cite{40_HBT_NA49} indicate that the
reaction dynamics do not change dramatically, therefore we do not expect drastic differences in the
rate. Indeed, the data at 40 AGeV look similar to the 160 AGeV case, and the calculated
rate, shown in figure \ref{Results_40}, also bears this similarity and achieves a good fit without tuning the setup of the model. Since the
initial temperature is lower and the QGP phase shorter in the 40 AGeV case, the partonic dilepton contribution is obviously much smaller, but nevertheless still present. Owing to the greater initial baryonic density, the in-medium
modifications of the vector mesons become more pronounced, most prominently visible in the
$\omega$ meson channel. Its downward mass shift drags the peak structure along the time evolution of the
fireball, creating a small bump on top of the completely dissolved $\rho$ meson that fills up the
low-mass region again. Its yield after freeze-out constitutes a visible
signal that may be experimentally observable with suitable energy resolution. The $\phi$ meson
contribution clearly sticks out above the smooth $\rho$ meson 'continuum'. To conclude, we find no
distinct differences in our calculation for the two beam energies probing dilepton production at SPS so
far, in accord with experimental findings. This indicates that the general setup of our model is fairly
robust. Future data at 20 and 80 AGeV will aid to test this statement.

\begin{figure}[!htb]
\begin{center}
\hspace*{-1.0cm}\epsfig{file=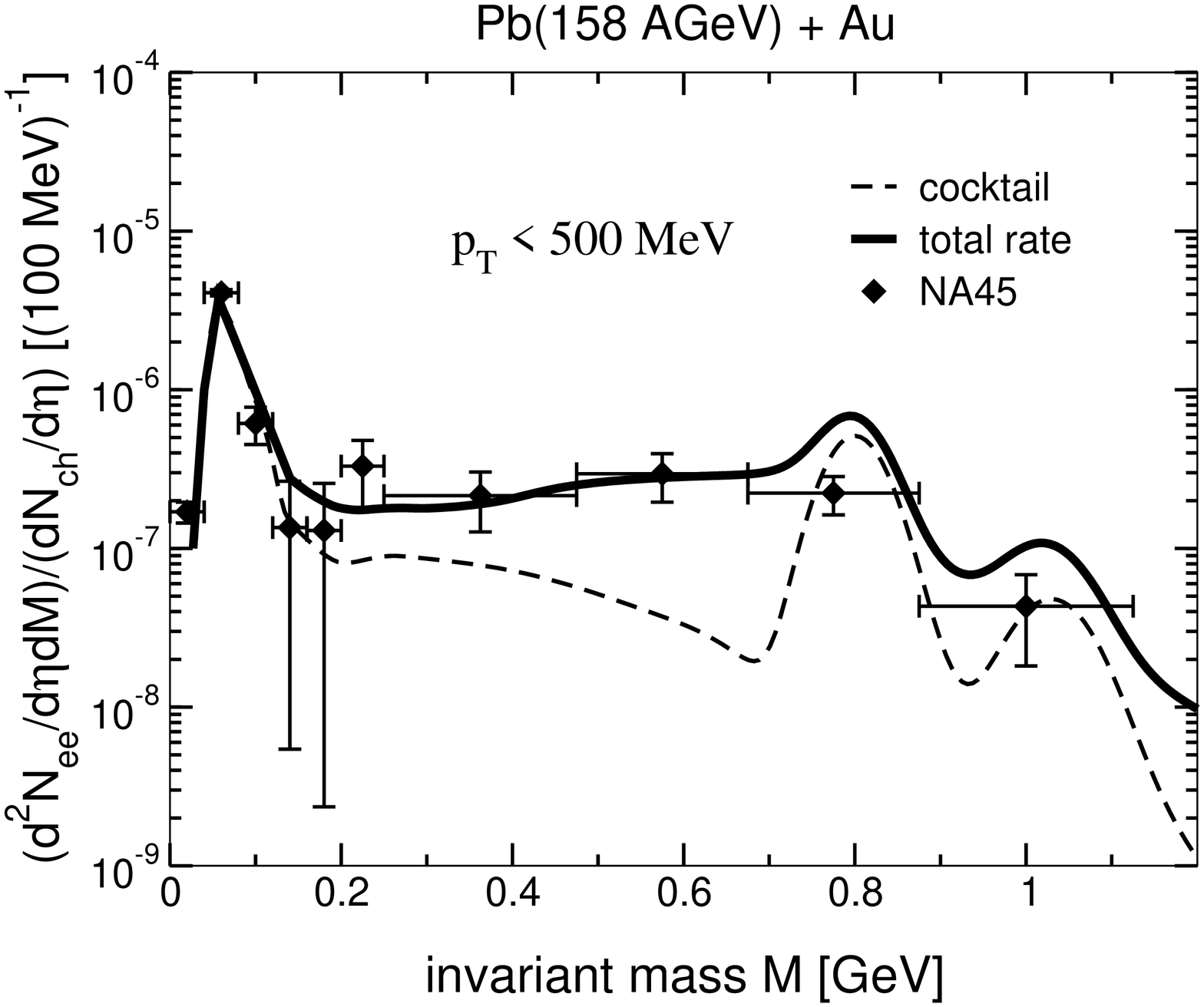, width=6.75cm, angle=-90} \hspace*{-1.0cm}\epsfig{file=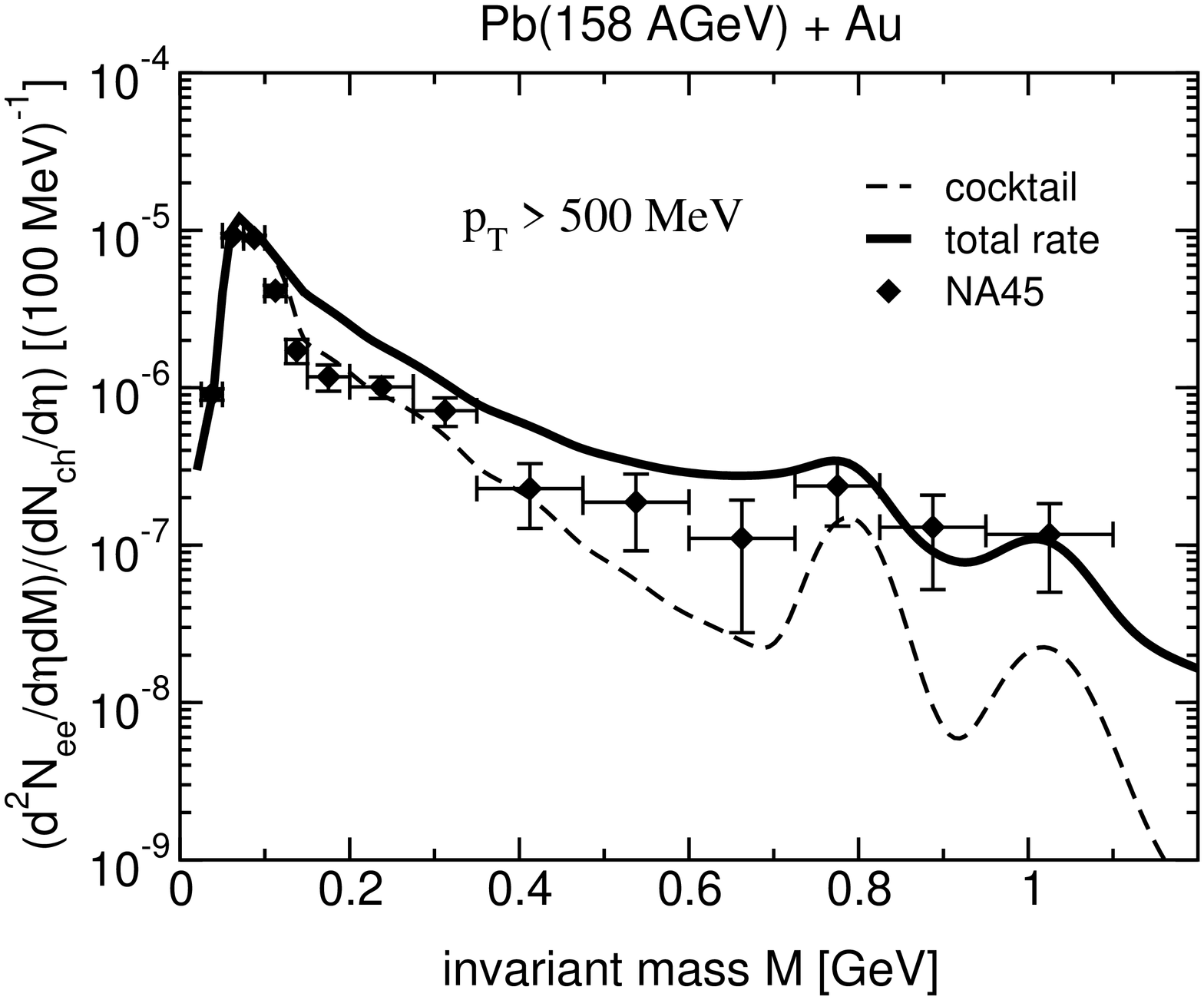,
width=6.75cm, angle=-90} \hspace*{-2.0cm}
\end{center}
\caption{\label{Results_160_PT}Left: Dilepton invariant mass spectra for transverse momenta of the $e^+
e^-$ pair ${\bf p}_t < 500$ MeV for the SPS CERES/NA45 Pb(158 AGeV)+Au experiment \cite{LENKEIT}. Shown are the data, the total rate and the cocktail contribution. Right: Same for
${\bf p}_t > 500$ MeV.}
\end{figure}

\begin{figure}[!htb]
\begin{center}
\hspace*{-1.0cm} \epsfig{file=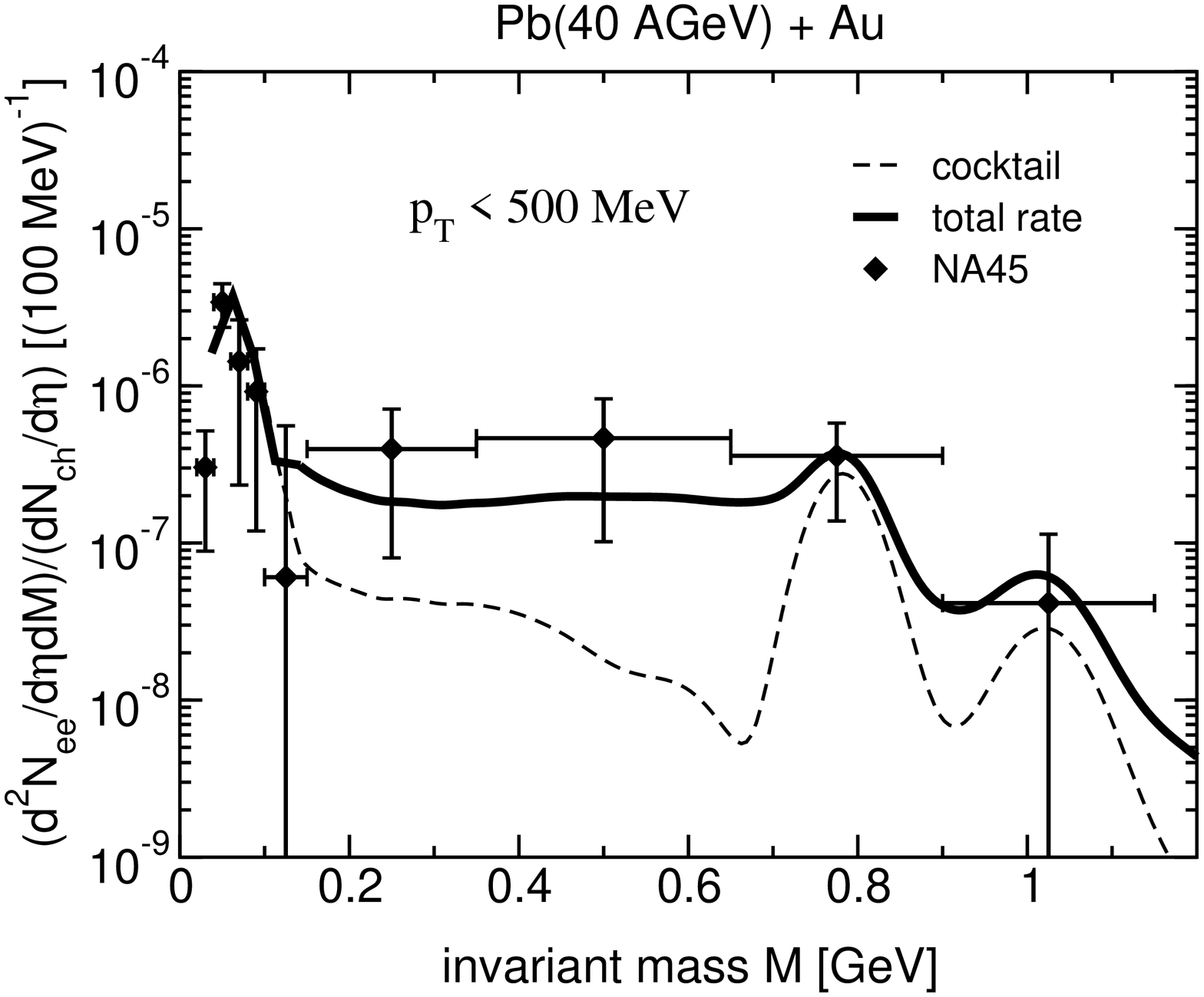, width=6.75cm, angle=-90}  \hspace*{-1.0cm} \epsfig{file=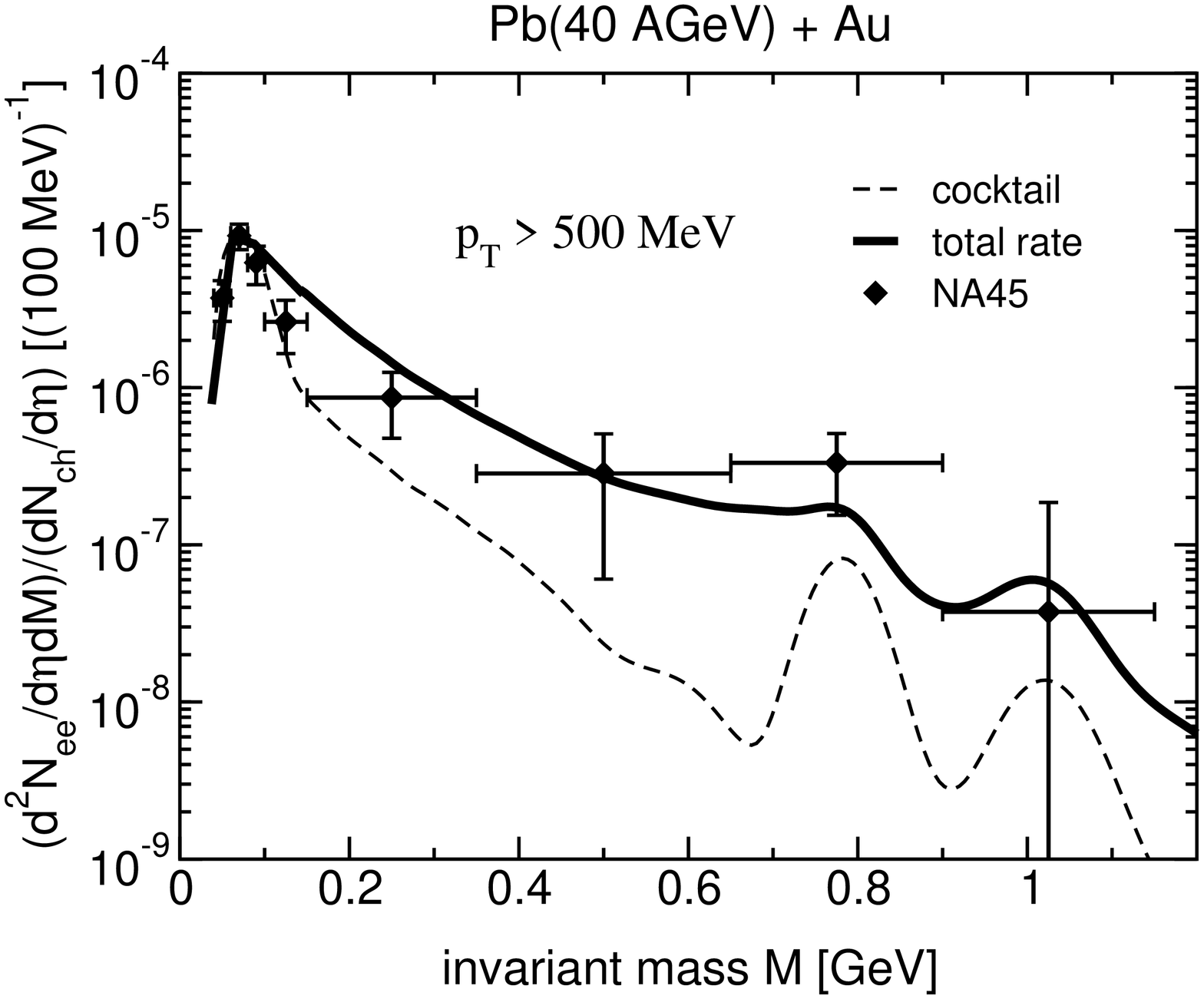,
width=6.75cm, angle=-90} \hspace*{-2.0cm}
\end{center}
\caption{\label{Results_40_PT}Left: Dilepton invariant mass spectra for transverse momenta of the $e^+
e^-$ pair ${\bf p}_t < 500$ MeV for the SPS CERES/NA45 Pb(40 AGeV)+Au experiment \cite{S_PRIV}. Shown are the data, the total rate and the cocktail contribution. Right: Same for
${\bf p}_t > 500$ MeV.}
\end{figure}

The fact that we moderately overestimate the data in the region between 200 and 300 MeV invariant
mass requires a comment. Since this range is dominated by the low-mass behaviour of the $\rho$ meson spectral function at
finite density, this behaviour may indicate that the influence of finite three-momentum on the spectra in that very region is non-negligible.
Consider the 160 AGeV and 40 AGeV data taken for different transverse momenta 
${\bf p}_t < 500$ MeV and $ {\bf p}_t > 500$ MeV, shown in figures~\ref{Results_160_PT} and \ref{Results_40_PT}. We observe that the general shape of the data pattern is
well described by the calculation for both ${\bf p}_t$ regions and both beam energies. However, for the high  ${\bf p}_t$
case, the calculation again overshoots the data in the low mass region whereas this is not so in the low
${\bf p}_t$ case. This can be traced to the use of the spectral functions for three-momentum equal to zero
in order to describe the photon spectral function in the hadronic phase. This approximation is of
limited validity at high ${\bf p}_t$ where the spectral function ought to become smaller than in our
approximation. Therefore we expect improved agreement with the data for the low invariant mass region
both at 40 and 160 AGeV once this effect is taken into account properly. 

\begin{figure}[!htb]
\begin{center}
\epsfig{file=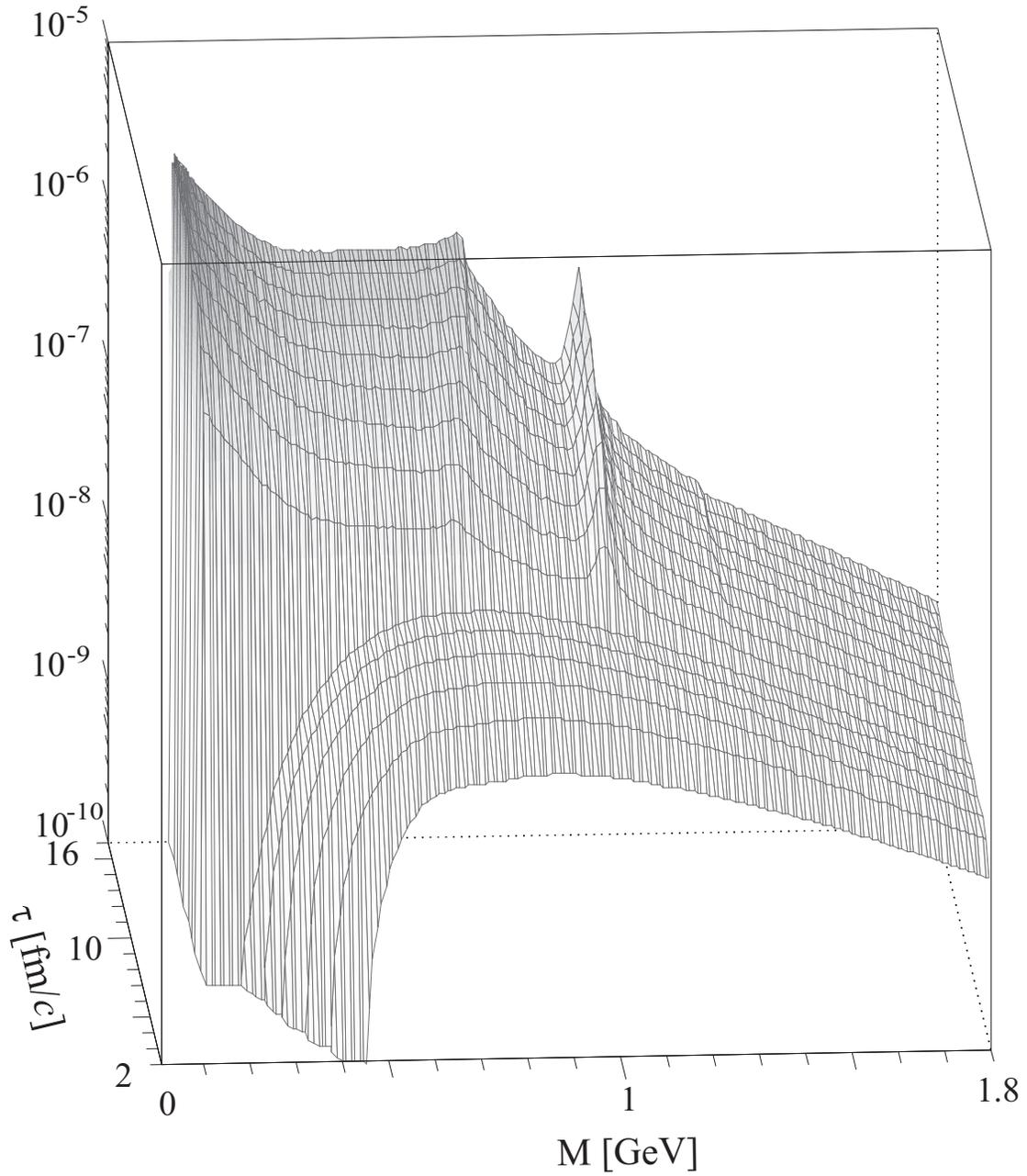, width=15cm}
\end{center}
\caption{\label{F-SPSTime}Time evolution of the (integrated) dilepton yield for the SPS scenario. The
vertical scale is the same as in fig.~\ref{Results_160}. The cocktail contribution has not been
included in this picture, and perfect resolution was assumed.
}
\end{figure}

Our setup of the fireball model enables us to gain additional detailed insight into the time evolution
of the dilepton yield. This is demonstrated in figure~\ref{F-SPSTime}. Here the different stages of the
fireball evolution leave distinct marks in the time-resolved dilepton yield.  For early times,
dileptons come entirely from the $q\overline{q}$ quasiparticle annihilation processes. The movement of
the invariant mass threshold reflects the temperature dependence of the quasiparticle mass which
decreases near the phase transition at $\tau \sim 7$ fm/c. One observes that, in spite of the growing
fireball volume, the contributions from later timeslices to the total yield become progressively less
important until the hadronic phase takes over at $\tau > 7 $ fm/c. This surprising behaviour is enforced by
the confinement factor $C(T)$ which reduces the thermodynamically active degrees of freedom
significantly near the phase boundary.
The system then enters the hadronic evolution phase without going through a mixed phase. The most
prominent feature is the rapid filling of the low invariant mass region through the density-broadened
$\rho$ meson spectrum which ends up as an enhanced pion continuum. The $\phi$ meson starts contributing
its characteristic peak and, as the system cools down further, the $\omega$ meson begins to emerge,
albeit as a broad structure. Note that while the hadronic contributions fill primarily the low
invariant mass region, their yield above 1 GeV is negligible in this late evolution phase. 

\subsection{RHIC at $\sqrt{s} = $ 200 AGeV}

For the RHIC scenario, thermally generated dileptons are dominant. Measurements of proton ratios
at $\sqrt{s} = $  130 AGeV indicate that the central collision region remains almost net-baryon free,
compared to SPS energies. Within a statistical thermal model, the particle ratios are accounted for by
a small baryon chemical potential of about 50 MeV at chemical freezeout \cite{RATIOS-RHIC}. At 200 AGeV, this
value is predicted to be even smaller. Effects of finite baryon density are therefore almost absent and
consequently both the $\rho$ and the $\phi$ are expected to show up in the spectrum as pronounced
structures, whereas there should be no clear trace of the in-medium $\omega$ due to its strong thermal
broadening. Contributions from Drell-Yan processes, which dominate in the very high invariant mass
region, are an order of magnitude smaller.

\begin{figure}[!htb]
\begin{center}
\hspace*{-1.0cm} \epsfig{file=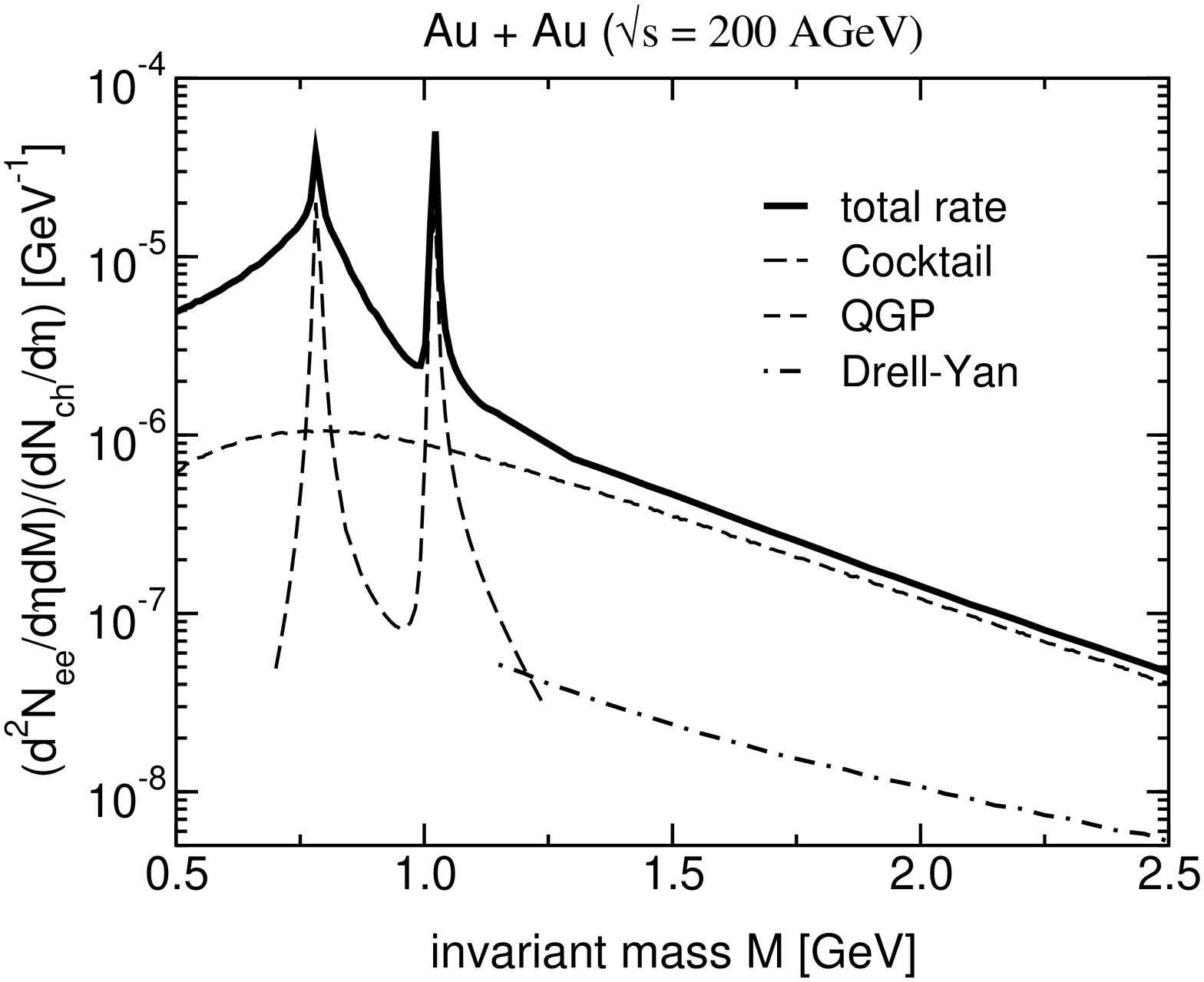, width=6.75cm, angle = -90} \hspace*{-1.0cm}
\epsfig{file=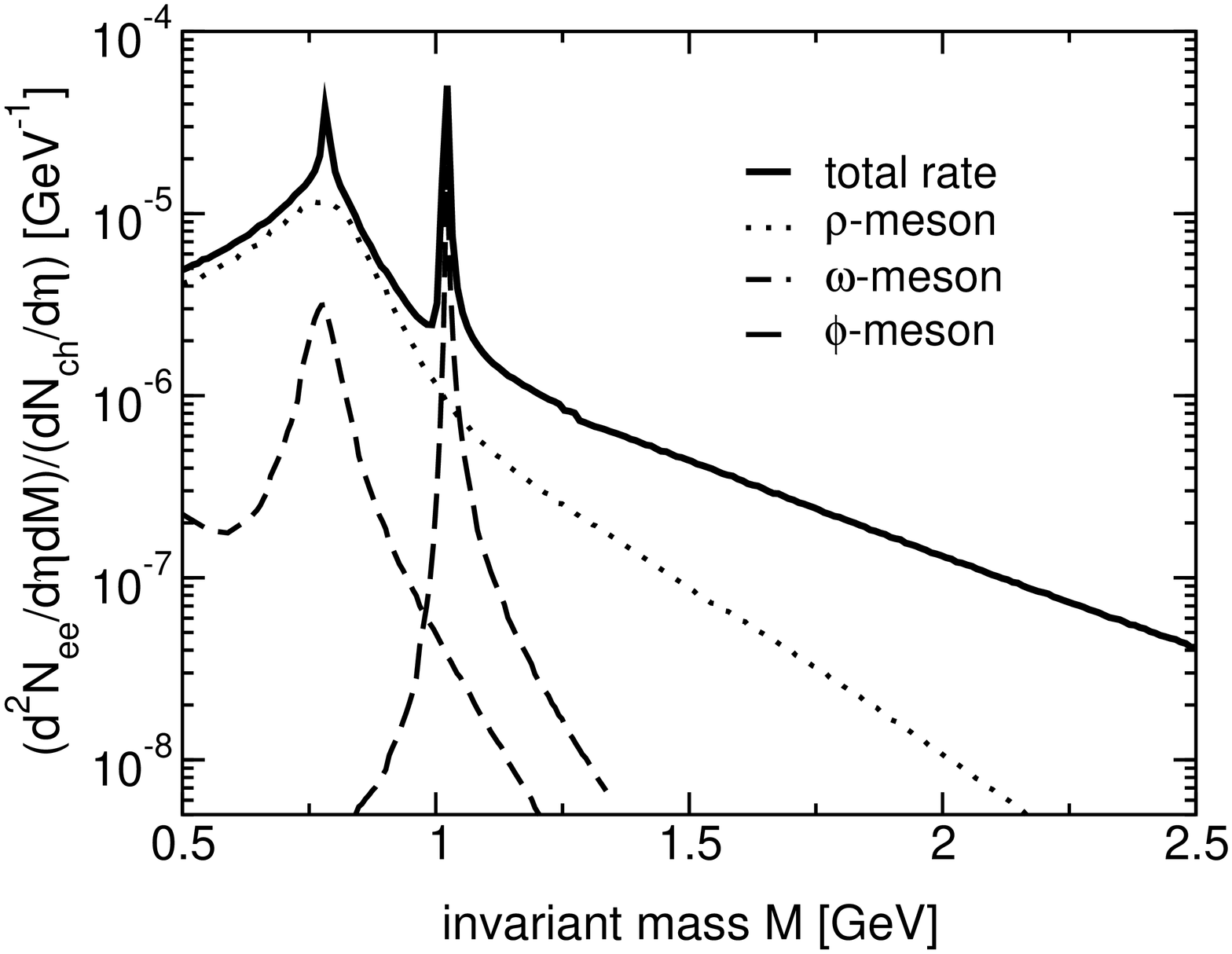, width=6.75cm, angle = -90} \hspace*{-2.0cm}
\end{center}
\caption{\label{Results_200}Left: Dilepton invariant mass spectra,
normalized to $dN_{ch}/d\eta = 650 $ \cite{dNdy_200}, in units of GeV$^{-1}$, for the RHIC experiment PHENIX at $\sqrt{s} = $ 200 AGeV. Shown are the total rate, the cocktail consisting of the after freeze-out decays of $\omega$ and $\phi$ vector mesons, the QGP contribution and the Drell-Yan yield. Right: Contributions from $\rho$-, $\omega$- and $\phi$-mesons (excluding after freeze-out yield) shown separately.}
\end{figure}

The corresponding prediction for the dilepton yield at 200 AGeV, including the schematic 
acceptance for the PHENIX detector, is shown in figure \ref{Results_200}. The $\omega$ and $\phi$ meson 
resonances clearly stick out over the smooth $\rho$ meson and QGP contributions. Although 
PHENIX will only start to measure at $M \geq$ 1 GeV, it may be possible to resolve the 
$\phi$ meson peak. However, a significant part of the peak strength is built up by the after  
freeze-out contributions, making it difficult to disentangle the in-medium modifications on 
the hadrons. 
For $M \geq 1.3$ GeV, the dilepton spectrum is dominated by thermal QGP radiation, outshining the hard Drell-Yan dilepton yield. 

Comparing our prediction for PHENIX with the one shown in ref.~\cite{RHIC},
we find rough agreement of the rate for the low mass region
below $\sim 1$ GeV. Although the dilepton yield from the QGP phase is suppressed in our case by the  factor $C(T)^2$ (cf. section \ref{subsec_qgp}), we still find an enhancement of a factor of about 4 in the range $1.3 - 2.5$ GeV over the rate in \cite{RHIC} that employs a (perturbative) chemical undersaturation model in the QGP phase. Owing to the non-perturbative nature of the QGP close to $T_C$, this model may not be applicable in that very region.  Note that our rate also constitutes only a lower limit there, so that the actual rate may even be larger. High precision data will allow to pin down one or the other model.

\subsection{Sensitivity to model parameters}

We would like to stress that the gross
features of our model are set, once the parametrization of the fireball
evolution has been matched to the hadronic observables and the EoS of
both phases has been constructed in accordance with lattice QCD and empirical constraints.
The remaining uncertainties, mainly about the initial state of the fireball, the
thermal masses of the quasiparticles and the detailed shape of the
EoS, do not alter the results substantially; they lead to only moderate
or even weak dependence on those parameters. Fine-tuning is still possible,
but only within the limits that retain consistency with the
overall framework.

\begin{figure}[!htb]
\begin{center}
\epsfig{file=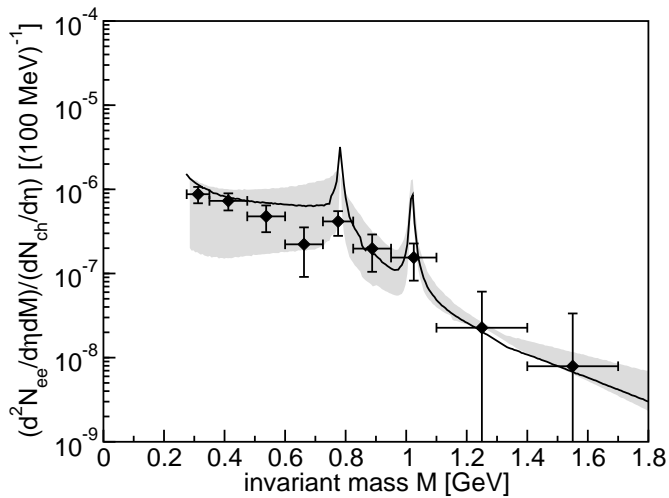, width=8cm, angle = -90}
\end{center}
\caption{\label{Errorband_160} Dilepton invariant mass spectra for the SPS CERES/NA45 experiment at 160 AGeV. Shown are data (symbols), the total rate introduced in  section \ref{subsec_sps} (solid line) and a band (shaded) that represents the range in the dilepton yield when varying the parameters $\tau_0$, $v_z^i$ and $T_c$. See text for details.}
\end{figure}

We have investigated the sensitivity of the model to parameter changes in some detail 
for the SPS
scenario at 160 AGeV. In order to get a theoretical 'error band', we investigated the 
extremes of our
parameter ranges, a combination of parameters that yields the largest and the smallest possible QGP contribution. The resulting range is shown in figure \ref{Errorband_160} as a grey 
band, together with the data points and the curve from the previous section.

Regarding the initial conditions, the largest uncertainty comes from the initial fireball formation length $z_0$ or, equivalently, the time $\tau_0$. We let it vary from 0.5 fm/$c$ to 2 fm/$c$, i.e. from fast to slow equilibration. Consider next the initial longitudinal flow velocity $v^i_z$. A variation of this
quantity from $0.3$ $c$ to $0.8$ $c$ implies a modification of the
pressure-driven accelerated motion in order to arrive at the same final velocity of
$0.9$ $c$, as determined by the rapidity distributions of the observed hadrons. 
Strong deviations from $v_z^i \simeq 0.5$ $c$, however, lead to inconsistent values for the thermal 
freeze-out temperature $T_{f}$. Finally, modifications of the critical temperature 
$T_C$ influence mainly the relative weight
of the contributions from the QGP phase and the
hadronic gas phase to the dilepton yield and hence change the shape of the dilepton spectra. Lattice data on $T_C$ suggest a range from 140 MeV for three massless, thermally active flavours to about 185 MeV for two massless flavours and a realistic pion mass.  Due to the self-consistent modelling of the temperature and volume evolution, there is no simple one-to-one correspondence of these parameters to the dilepton yield.  

The upper limit of the grey band in figure \ref{Errorband_160} now corresponds to the scenario 
with a {\em small} QGP contribution, \emph{i.e.} small $\tau_0$, large $v_z^i$ and high $T_C$. On the 
other hand, the lower limit includes a {\em large} QGP contribution with large $\tau_0$, small 
$v_z^i$ and low $T_c$. It is instructive to note that the shape of the spectrum changes only 
moderately above 1 GeV invariant mass within these extreme parameter variations. The first scenario, 
however, tends to overestimate the data in the region of the $\rho$ peak, whereas the second 
scenario does not leave enough time for the hadronic phase to build up the $e^+e^-$ excess in 
the low-mass region between 200 and 800 MeV, effectively ruling out a large QGP contribution.

\section{Concluding Remarks}

We have calculated dilepton radiation from an expanding fireball created in ultra-relativistic 
heavy-ion collisions over a wide range of beam energies, from SPS 40 AGeV via SPS 160 AGeV to 
RHIC $\sqrt{s} = $ 200 AGeV.

We have explored the evolution of a fireball through the quark-gluon and
hadronic phases as they are assumed to be characteristic
of the QCD equation of state. Furthermore, we have emphasized the relevance of using input and
constraints from lattice QCD thermodynamics in the analysis of dilepton signals
from the expanding fireball. The fact that the measured dilepton rates
could be reproduced using a fireball model fixed by hadronic information
only is a nontrivial confirmation of its validity. Further data for different
energies and/or nuclear systems will help to strengthen or weaken this
position.

High resolution measurements of the invariant mass spectrum around the
region of the $\omega$ and the $\phi$ mass can reveal information on the
average density and temperature in the hadronic phase. The
broadening and mass shift of the $\omega$ meson with temperature can serve as a 'thermometer' once 
the cocktail contribution is reliably assessed. A visible enhancement in the $\omega$ 
region beyond the rate originating from $\omega$-decays after
freeze-out would hint at a temperature
which is on average lower than assumed in this model and therefore
point to a lower freeze-out temperature or alternative in-medium effects on hadrons. 
On the other hand, the
$\phi$ meson signal, being nearly unmodified by temperature effects over a comparably long timespan, 
may be used as a 'standard candle' at RHIC conditions.

Detailed knowledge of the slope of the invariant mass spectrum above 1 GeV
gives an indication of the average temperature in the quark-gluon phase, serving as a constraint 
for the fireball dynamics in the QGP region. Its total normalization is useful to estimate 
the effects of a possible underpopulation of the quark and gluon
phase space in early stages of the evolution, or of phenomenological models of the QGP,
such as the confinement model presented in section \ref{sec_quasiparticles}.
It is remarkable that this confinement model is able to reproduce the
data despite the suppression of dilepton radiation near the phase
transition temperature.

Measurements using different nuclear systems or different impact parameter can provide different relative
weights of QGP and hadronic dilepton radiation to the observed yield. One may thus hope to 
disentangle these contributions and
test the model assumptions above and below the phase transition separately,
at least qualitatively.

The present scenario supports the hypothesis that the quark-gluon
phase is actually reached, at a transient stage, in heavy-ion
collisions at CERN/SPS and (more pronouncedly) at RHIC. Within our
outlined approach, a
purely hadronic framework would not be successful in comparison
with existing data. 
The reason is that the relative strength of
hadronic contributions to the high mass region and the low
mass region is quite different as compared to those of the
QGP (cf. figure~\ref{Results_160}). Purely hadronic scenarios which are
able to account for the low mass region would necessarily
fail in the high mass region and vice versa.
Furthermore, fireball thermodynamics as described in this work leads to temperatures
well above $T_C$ for a broad range of initial conditions, though not
necessarily to significant contributions to the dilepton yield from the high-temperature region.
The upcoming measurements at RHIC can be expected to provide further 
interesting insights.

\section*{Acknowledgements}

We are grateful to Ralf Rapp, Jochen Wambach, Francois Arleo, Alberto Polleri,
Hans J. Specht and Sanja Damjanovic for
stimulating discussions and exchanges in the course of preparing this paper.

\appendix

\section{Fireball properties}

In the following, we summarize key properties of the different fireball scenarios discussed in the
paper. We characterize each scenario by the proper time for fireball formation, $\tau_0$, the phase
transition time $\tau_c$ and the freeze-out time $\tau_f$. Note that the corresponding times in the
c.m. frame can be considerably larger, especially for RHIC conditions where volume elements are
travelling near the speed of light at the fireball edge. Additionally, we quote the volume at
freeze-out and the flow velocities at rms radius $v_\perp^f$ and maximum longitudinal extention
$v_z^f$. In order to compare the thermodynamic conditions, we provide $s/\rho$, the entropy per baryon.

\vspace{1cm}

\begin{table}[htb]
\begin{tabular}{|l|c|c|c|c|c|c|c|c|}
\hline
&$\tau_0$ [fm/$c$] & $\tau_c$ [fm/$c$] & $\tau_f$ [fm/$c$]&$V_f$ [fm$^3$]&
$v_\perp^f$& $v_z^i$&  $v_z^f$&$s/\rho_B$\\
\hline
\hline
SPS 40 AGeV& 1.5&4.0&15.0&7040&0.36& 0.45  &0.75&13\\
\hline
SPS 160 AGeV& 1.0&6.5&16.0&14.400&0.53&0.45  &0.9&26\\
\hline
RHIC 200 AGeV& 0.6&10.0 &18.0 &99846.1 &0.56& 0.9 &0.9985 &250\\
\hline
\end{tabular}
\caption{\label{T-Fireball} Key parameters of the fireball evolution scenarios} 
\end{table}

\vspace{1cm}

Note that the expansion is modelled in proper time. The lifetime
of the fireball in the RHIC c.m. frame, for example, is more than four times
larger due to time dilatation at the fireball edge.


\end{document}